\documentclass[10pt,letterpaper,final,twocolumn,journal]{IEEEtran}
\usepackage{setspace}
\usepackage{nomencl}
\makenomenclature
\usepackage{graphicx}

\usepackage{soul}

\usepackage{algorithm}
\usepackage{algorithmic}
\usepackage{epstopdf}
\usepackage{epsfig}
\usepackage[cmex10]{amsmath}
\usepackage{cite}
\usepackage{amssymb}
\usepackage[usenames,dvipsnames]{color} 
\usepackage{multirow}
\usepackage{array}
\long\def\symbolfootnote[#1]#2{\begingroup\def\thefootnote{\fnsymbol{footnote}}\footnote[#1]{#2}\endgroup}
\usepackage[table]{xcolor}
\usepackage{tabularx}
\usepackage{chngcntr}
\usepackage{booktabs}
\usepackage{caption}
\usepackage{subcaption}
\usepackage{float}
\usepackage{mathtools,cuted}
\usepackage{braket}

\usepackage{hyperref}

\usepackage{amsthm}
\theoremstyle{definition}
\newtheorem{definition}{Definition}
\newtheorem{proposition}{Proposition}
\newtheorem{corollary}{Corollary}
\newtheorem*{remark}{Remark}

\newcommand{\subparagraph}{}
\newcommand{\eqdef}{\stackrel{\triangle}{=}}
\usepackage{braket}

\begin{document}

\markboth{IEEE Transactions on Wireless Communications}{Accepted paper}

\title{{The Entanglement-Assisted Communication Capacity over Quantum Trajectories}}
\author{Daryus~Chandra,~\IEEEmembership{Member,~IEEE}, Marcello~Caleffi,~\IEEEmembership{Senior Member,~IEEE}, Angela~Sara~Cacciapuoti,~\IEEEmembership{Senior Member,~IEEE}
	\thanks{D. Chandra, is with the School of Electronics and Computer Science, University of Southampton, Southampton, SO17 1BJ, United Kingdom. E-mail: \href{mailto:daryus.chandra@soton.ac.uk}{daryus.chandra@soton.ac.uk}.}
	\thanks{M. Caleffi and A.S. Cacciapuoti are with the Department of Electrical Engineering and Information Technology, University of Naples Federico II, Naples, 80125, Italy. E-mail: \href{mailto:marcello.caleffi@unina.it}{marcello.caleffi@unina.it}, \href{mailto:angelasara.cacciapuoti@unina.it}{angelasara.cacciapuoti@unina.it}. Web: \href{http://www.quantuminternet.it}{www.quantuminternet.it}.}
	\thanks{M. Caleffi and A.S. Cacciapuoti are also with the Laboratorio Nazionale di Comunicazioni Multimediali, National Inter-University Consortium for Telecommunications (CNIT), Naples, 80126, Italy.}
	\thanks{A.S Cacciapuoti, M. Caleffi and D. Chandra would like to acknowledge the financial support of the project ``Towards the Quantum Internet: A Multidisciplinary Effort'', University of Naples Federico II. A.S Cacciapuoti and M. Caleffi would like to to acknowledge as well the financial support of PON project ``S4E - Sistemi di Sicurezza e Protezione per l’Ambiente Mare''.}
}

\maketitle

\begin{abstract} 
The unique and often-weird properties of quantum mechanics allow an information carrier to propagate through multiple trajectories of quantum channels simultaneously. This ultimately leads us to quantum trajectories with an indefinite causal order of quantum channels. It has been shown that indefinite causal order enables the violation of \textit{bottleneck capacity}, which bounds the amount of the transferable classical and quantum information through a classical trajectory with a well-defined causal order of quantum channels. In this treatise, we investigate this beneficial property in the realm of both entanglement-assisted classical and quantum communications. To this aim, we derive closed-form capacity expressions of entanglement-assisted classical and quantum communication for arbitrary quantum Pauli channels over classical and quantum trajectories. We show that by exploiting the indefinite causal order of quantum channels, we obtain capacity gains over classical trajectory as well as the violation of bottleneck capacity for various practical scenarios. Furthermore, we determine the operating region where entanglement-assisted communication over quantum trajectory obtains capacity gain against classical trajectory and where the entanglement-assisted communication over quantum trajectory violates the bottleneck capacity.
\end{abstract}


\pagenumbering{arabic}

\section{Introduction}
\label{Introduction}

Classical information theory has been revitalized into the quantum realm, where it is formally known as quantum Shannon theory or quantum information theory~\cite{bennett1998quantum}. In both classical and quantum information theory framework, the trajectory traversed by the information carrier is causally well-defined. Interestingly, due to the property of quantum information, the information carrier can traverse quantum channels in a superposition of multiple trajectories, characterized by different causal orders. This implies that the path traversed by the information carrier exhibits an indefinite causal order of quantum channels~\cite{oreshkov2019time}. Consequently, there is a novel paradigm of quantum information theory involving the existence of indefinite causal order of quantum channels, which leads to a new frontier research field~\cite{chiribella2019quantum}.

\begin{figure}[t]
\center
\includegraphics[width=\linewidth]{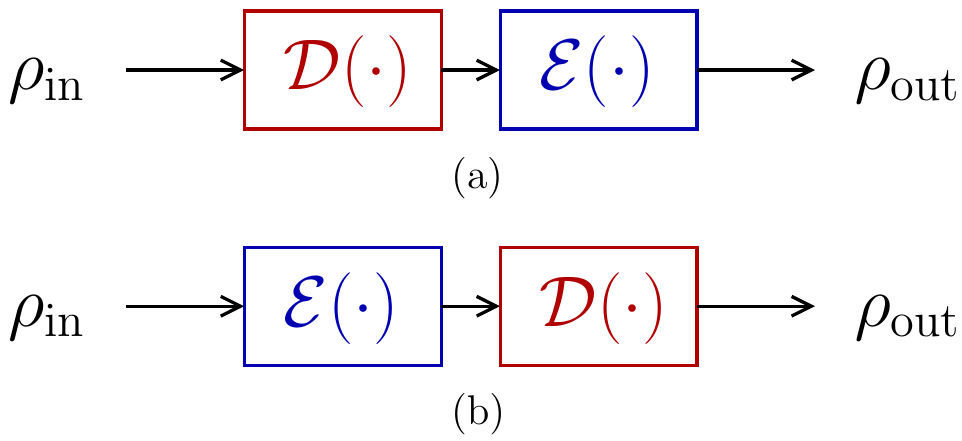}
\caption{An illustration of an information carrier traverses two different classical trajectories with a definite causal order. (a) The information carrier traverses quantum channels $\mathcal{D}(\cdot) \rightarrow \mathcal{E}(\cdot)$. (b) The information carrier traverses quantum channels $\mathcal{E}(\cdot) \rightarrow \mathcal{D}(\cdot)$.}
\label{fig:serial}
\end{figure}

For instance, let us observe Fig.~\ref{fig:serial}. Assume that in order to transfer the information from the source to the destination, we have to utilize two quantum channels denoted by $\mathcal{D}(\cdot)$ and $\mathcal{E}(\cdot)$. In the conventional framework of quantum information theory, the information carrier can traverse through the quantum channel $\mathcal{D}(\cdot)$ first and then followed by quantum channel $\mathcal{E}(\cdot)$ or through the quantum channel $\mathcal{E}(\cdot)$ first and then followed by quantum channel $\mathcal{D}(\cdot)$. In both cases, the causal order of the quantum channels are well-defined, it is either $\mathcal{D}(\cdot) \rightarrow \mathcal{E}(\cdot)$ or $\mathcal{E}(\cdot) \rightarrow \mathcal{D}(\cdot)$. In other words, it is said that they have a \textit{definite causal order}. Furthermore, we may refer to the path traversed by information carrier with a definite causal order as \textit{classical trajectory}.

\begin{figure}[t]
\center
\includegraphics[width=\linewidth]{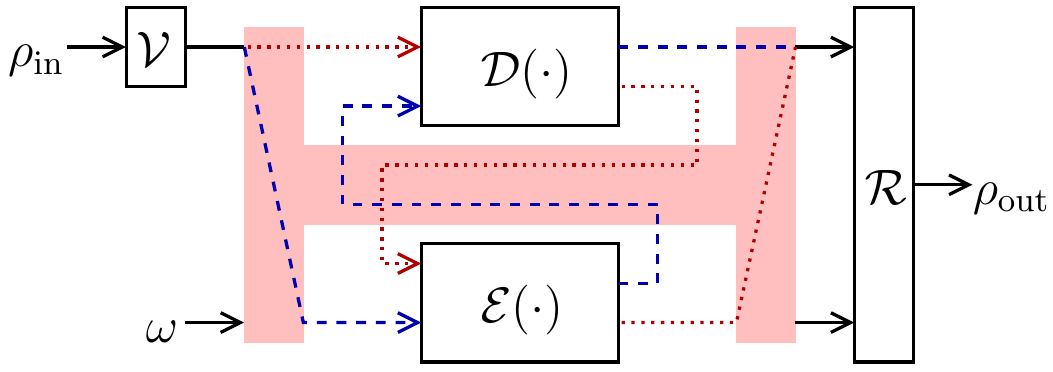}
\caption{An illustration of a quantum trajectory traversed by information carrier. The superposition of classical trajectories may be determined using the control qubit $\omega$. The result is a supermap $\mathcal{S}_{\omega}(\mathcal{E},\mathcal{D})$ denoted in pink, which constitutes the superposition of two causal orders, $\mathcal{D}(\cdot) \rightarrow \mathcal{E}(\cdot)$ denoted by red dotted lines and $\mathcal{E}(\cdot) \rightarrow \mathcal{D}(\cdot)$ denoted by blue dashed lines. Within the figure, $\mathcal{V}$ and $\mathcal{R}$ denote a possible encoding-decoding pair. The whole operation will be elaborated further in Section~\ref{The Classical and Quantum Trajectory}.}
\label{fig:switch}
\end{figure}

Interestingly, the properties of quantum information allow the information carrier to traverse the quantum channels in the superposition of both causal orders as illustrated in Fig.~\ref{fig:switch}. Precisely, in order to transfer the information from the source to the destination, the information carrier can traverse both possible combinations of the classical trajectories of the quantum channels simultaneously, i.e. $\mathcal{E}(\cdot) \rightarrow \mathcal{D}(\cdot)$ and $\mathcal{D}(\cdot) \rightarrow \mathcal{E}(\cdot)$. Consequently, the superposition of both classical trajectories traversed by the information carrier exhibits an \textit{indefinite causal order}. We may refer to the path traversed by an information carrier with an indefinite causal order as \textit{quantum trajectory}. 

\begin{remark}
	The term \textit{trajectory} is used in the relevant literature to denote the path -- generally assumed being constituted by a sequence of quantum channels -- traversed by the information carrier. If the quantum channels are traversed in a well-defined causal order, the information carrier is said to propagate through a \textit{classical trajectory}. Conversely, whenever the order of the quantum channels cannot be expressed as a well-defined causal order, the information carrier propagates through a \textit{quantum trajectory}. In other words, quantum mechanics allows the information carrier to traverse quantum channels placed in a quantum configuration.
\end{remark}

As counter-intuitive as it seems, the ability of information carrier traversing a superposition of classical trajectories has been experimentally verified using a quantum device called \textit{quantum switch}~\cite{procopio2015experimental, rubino2017experimental, goswami2018indefinite, rubino2019experimental, guo2020experimental, taddei2020experimental, goswami2020experiments, goswami2020increasing, rubino2021experimental}. Until recently, the alluring benefits of the indefinite casual order of quantum process have been reported for various aspects of quantum information processing, including quantum computation~\cite{chiribella2013quantum, araujo2014computational}, noiseless quantum teleportation~\cite{caleffi2020quantum}, communication complexity~\cite{guerin2016exponential}, quantum resource theory~\cite{feix2015quantum, kristjansson2020resource}, quantum metrology~\cite{zhao2020quantum}, discrimination of quantum process~\cite{koudia2019superposition}, and ultimately for boosting the channel capacities of quantum and classical communications over quantum channels~\cite{salek2018quantum, ebler2018enhanced, cacciapuoti2019capacity, chiribella2021indefinite}.

From a communication engineering point of view, the pivotal question is always how to quantify the advantage that can be obtained from exploiting the indefinite causal order of quantum channels for enhancing the quality of classical and quantum communication. In this treatise, we aim for answering the remaining open questions on how much advantage we can glean from the indefinite causal order of quantum channels to improve the capacities of both entanglement-assisted classical and quantum communication. More specifically, within the Quantum Internet framework, multiple quantum devices are interconnected via pre-shared entanglement for facilitating various applications that require the exchange of classical and quantum information amongst the quantum devices, including quantum communications~\cite{chiani2020piggybacking, guerrini2020quantum}, quantum cryptography~\cite{pirandola2020advances}, quantum sensing~\cite{paris2009quantum, degen2017quantum}, distributed quantum computation~\cite{van2016path, cuomo2020towards}, blind quantum computation~\cite{broadbent2009universal, fitzsimons2017private}, quantum-secure direct-communication (QSDC)~\cite{deng2003two, deng2004secure, wang2005quantum, sun2020toward}, and quantum-secure secret-sharing~\cite{gottesman2000theory}. Therefore, the pre-shared entanglement can be viewed as the primary consumable resources for enabling entanglement-assisted classical and quantum communications within the Quantum Internet framework~\cite{kimble2008quantum, wehner2018quantum, caleffi2018quantum, caleffi2020rise, cuomo2020towards}. Ultimately, the advantage gleaned from the indefinite causal order of quantum channels for entanglement-assisted classical and quantum communication can be immediately extended to the aforementioned applications. Thus, the analysis of entanglement-assisted communication over quantum trajectory will provide a critical milestone for the development of Quantum Internet. 

In this treatise, we consider quantum superdense coding protocol~\cite{bennett1992communication} as our model for entanglement-assisted classical communication since a single-letter capacity formulation can be derived for quantum Pauli channel~\cite{bennett1999entanglement}. More specifically, quantum Pauli channel constitutes a set of quantum channel models with various practical applications. Additionally, an entanglement-assisted classical communication constituted by quantum superdense coding is known to be the optimal scheme of utilizing a single use of quantum channel and a pair of pre-shared maximally-entangled quantum state in exchange for two classical bits~\cite{bennett1992communication}. Furthermore, the quantum superdense coding versus quantum teleportation trade-off suggests that the capacity formulation of entanglement-assisted quantum communication can be obtained directly from their classical counterparts~\cite{bowen2002entanglement, hsieh2010trading}.

Against this background, our contributions can be summarized as follows:
\begin{itemize}
\item \textit{We derive the general formulation of entanglement-assisted classical communication capacity over quantum trajectory for various scenarios involving quantum Pauli channels}.
\item \textit{We determine the operating region where entanglement-assisted communication over quantum trajectory obtains capacity gain against classical trajectory. Additionally, we also portray the operating region where entanglement-assisted communication over quantum trajectory violates the bottleneck capacity, which represents stringent upper-bound of communication capacity over definite causal order}.
\item \textit{We present the achievable capacity of entanglement-assisted quantum communication over quantum trajectory, which is obtained via quantum superdense versus quantum teleportation trade-off}.
\end{itemize}

The rest of this treatise is organized as follows. We provide a comprehensive comparison between our work and the state-of-the-art in Section~\ref{Related Works}. We present the quantum channel models considered as well as a brief description of quantum superdense coding in Section~\ref{Preliminaries}. Furthermore, we also provide the tools required for evaluating the capacity of entanglement-assisted classical communication based on the description of quantum channel models and quantum superdense coding. It is followed by Section~\ref{The Classical and Quantum Trajectory} where we detail the formal desription of the classical and quantum trajectory. The main results of the entanglement-assisted classical and quantum communication capacity are presented in Section~\ref{The Entanglement-Assisted Capacity over Classical and Quantum Trajectory}. Finally, we conclude our work in Section~\ref{Conclusions and Future Works} by also providing several potential directions for future research.

\section{Related Works}
\label{Related Works}

The first demonstration of the beneficial capacity gains obtained from the indefinite causal order of quantum channels was presented in~\cite{salek2018quantum}, which marks the lower bound of quantum communication capacity over quantum trajectory. More specifically, this lower bound is represented by unassisted quantum communication capacity obtained based on the entropy measure of the quantum channels~\cite{lloyd1997capacity, devetak2005private}. By contrast, quantum communication capacity is upper-bounded by two-way entanglement-assisted quantum communication capacity, whose formulation is calculated via relative entropy entanglement of the Choi matrix~\cite{pirandola2017fundamental}. Relying on this formulation, the upper-bound of quantum communication capacity over quantum trajectory has been recently presented in~\cite{cacciapuoti2019capacity}. Additionally, the lower-bound of quantum communication capacity given in~\cite{cacciapuoti2019capacity} is tighter than that in~\cite{salek2018quantum}. Hence, the operating capacity of quantum communication over quantum trajectory has been established. However, a noticeable gap can be observed between these lower- and upper- bounds, which may cause an inconvenience for determining the exact capacity for specific applications, including those of entanglement-assisted-based described in Section~\ref{Introduction}. Thus, to navigate the concept of quantum trajectory closer to practical purposes, in this treatise, we consider the one-way entanglement-assisted -- both classical and quantum -- communication in our investigation, whose capacity conceivably lies between these lower- and upper- bounds\footnote{The term \textit{unassisted} refers to a scenario where the source and the destination does not share a pre-shared EPR pair, while the term \textit{entanglement-assisted} assumes that the source and the destination have pre-shared maximally-entangled quantum state such as EPR pairs. The term \textit{one-way} refers to a scenario where classical communications can be conducted in one direction only, i.e. from the source to the destination (forward direction), while the term \textit{two-way} refers to a scenario where classical communications can be performed in both directions, i.e. from the source to the destination and vice versa (forward and backward directions). Entanglement-assisted communication exhibits higher capacity compared to the unassisted one and the two-way entanglement-assisted communication may attain higher capacity than the one-way one.}.

Since a quantum channel can be used for transferring both classical and quantum information, the investigation related to indefinite causal order of quantum channel is extended to the world of classical communication. However, currently, there is no closed-form formula for determining the exact classical communication capacity over a quantum channel, except for the widely known Holevo bound~\cite{holevo1973bounds, schumacher1997sending}, which marks the upper-bound of accessible classical information of the unassisted classical communication over a quantum channel. Consequently, it is also generally hard to find a single-letter formula for establishing the classical communication capacity over an indefinite causal order of quantum channels for a wide range channel parameters. However, for very specific cases, such as fully-depolarizing quantum channel and quantum entanglement-breaking channel, the unassisted classical communication capacities over quantum trajectory have been determined~\cite{ebler2018enhanced, abbott2020communication}. On the other hand, an entanglement-assisted classical communication constituted by quantum superdense coding is known to be the optimal scheme of utilizing a single use of quantum channel and a pair of pre-shared maximally-entangled quantum state in exchange for two classical bits~\cite{bennett1992communication}. Therefore, a single-letter formula of entanglement-assisted classical communication capacity can be derived for a wide range quantum channel parameters~\cite{bennett1999entanglement}. Relying on these facts, ultimately, we derive closed-form expressions for entanglement-assisted classical communication capacity over quantum trajectories.

\section{Preliminaries}
\label{Preliminaries}

In the classical domain, the information is conveyed by binary digit (bit), which can carry a value of ``0'' or ``1'' at a given time. By contrast, the information in quantum domain is carried by the quantum bit (qubit), where it can be used to represent ``0'' or ``1'' or even the superposition of both values. Thus, a qubit can also be used to carry classical information if the qubit is wisely encoded. Similar to the classical domain, the transfer of quantum information between the source and the destination is affected by the noise characterized by the quantum channel $\mathcal{N}(\cdot)$. However, differently from the classical domain, for a given quantum channel $\mathcal{N}(\cdot)$, we may have several different notions of communication capacity -- the maximal amount of information may be transferred reliably under a certain encoding and decoding procedure -- which include quantum communication capacity $Q$ as well as classical communication capacity $C$. More precisely, the quantum communication capacity $Q$ quantifies the maximum amount of quantum information that can be communicated from the source to the destination over many independent uses of a quantum channel $\mathcal{N}(\cdot)$. Similarly, the classical communication capacity $C$ quantifies the amount of classical information that can be reliably transferred over many independent uses of a quantum channel $\mathcal{N}(\cdot)$. However, in this case, the information is encapsulated using a carefully selected classical-to-quantum mapping. We refer the readers to~\cite{gyongyosi2018survey} and~\cite{koudia2021deep} for an in-depth overview about classical and quantum communication capacities. In this treatise, we consider the so-called entanglement-assisted communication capacities, which differ from unassisted communication capacities, since maximally-entangled quantum states are pre-shared between the source and the destination before the communication is commenced. In this section, we provide the fundamental background required for establishing the main results of our paper.

\subsection{Quantum Channel}
\label{Quantum Channel}

An arbitrary pure quantum state, denoted by $\ket{\psi}$, can be formally expressed as a superposition of the orthogonal basis states $\{ \ket{i} \}$ as follows:
\begin{equation}
\ket{\psi} = \sum_{i} \alpha_i \ket{i}.
\end{equation}
The measurement of the quantum state $\ket{\psi}$ in the orthogonal basis collapses the superposition of the quantum state into one of the basis states, say $\ket{i}$, with probability $|\alpha_i|^2$. Hence, the coefficients $\{ \alpha_i \}$ are subjected to the normalization condition $\sum_{i} |\alpha_i|^2 = 1$. A mixed state, i.e., the statistical mixture of multiple pure quantum states $\{\ket{\psi_i}\}$, can be described using the so-called density matrix $\rho$, which is defined as
\begin{equation}
\rho \eqdef \sum_{i} p_i \ket{\psi_i} \bra{\psi_i},
\end{equation}
with $\ket{\cdot}\bra{\cdot}$ denoting the outer product between two pure quantum states.

\begin{definition} \textit{Quantum channel}~\cite{nielsen2010quantum}. A quantum channel $\mathcal{N}(\cdot)$ is a completely-positive trace-preserving (CPTP) map acting on arbitrary quantum states, which can be written in the \textit{operator-sum representation} as follows:
\begin{equation}
\mathcal{N}(\rho) = \sum_{i} N_i \rho N_i^{\dagger},
\label{eq:qchannel}
\end{equation}
where $\lbrace N_i \rbrace$ denotes a set of operators -- referred to as \textit{Kraus operators} -- satisfying the following completeness criterion:
\begin{equation}
\sum_{i} N_i^{\dagger} N_i = I.
\label{eq:completeness}
\end{equation}
\end{definition}

A special class of the quantum channel representation of~\eqref{eq:qchannel} is constituted by a quantum channel where the Kraus operators can be expressed in terms of unitary operators $\lbrace U_i \rbrace$ as follows: 
\begin{equation}
\mathcal{N}(\rho) = \sum_{i} p_i U_i \rho U_i^{\dagger},
\label{eq:sum‐unitary}
\end{equation}
where $U_i$ is a unitary operator satisfying $U_i^{\dagger}U_i = I$ and $\{ p_i \}_i$ is a probability distribution satisfying $\sum_{i} p_i = 1$. Hence, the relationship between the Kraus operators and the unitary operators is given by $N_i = \sqrt{p_i}U_i$.

\begin{definition} \textit{Quantum Pauli channel}~\cite{nielsen2010quantum}. A quantum channel is referred to as quantum Pauli channel based on the following map:
\begin{equation}
\mathcal{N}(\rho) = \sum_{i = 0}^{3} p_i \sigma_i \rho \sigma_i^{\dagger},
\label{eq:quantum‐pauli‐channel}
\end{equation}
where $\sigma_0 = I$, $\sigma_1 = X$, $\sigma_2 = Y$, and $\sigma_3 = Z$ are the Pauli matrices\footnote{In the rest of this treatise, we remove the ${\dagger}$ of~\eqref{eq:quantum‐pauli‐channel} for quantum Pauli channels since Pauli matrices are Hermitian.}.
\end{definition}

Let us now consider a joint quantum state $\rho_{AB}$ shared between two parties $A$ and $B$, where quantum channel $\mathcal{N}_A(\cdot)$ affects only the quantum state at $A$. Since we consider that the quantum state at $B$ undergoes an ``error-free'' identity channel, the extended mapping of $\mathcal{N}_A(\cdot)$ on the joint quantum state $\rho_{AB}$ is given by
\begin{equation}
(\mathcal{N}_A \otimes I_B)(\rho_{AB}) = \sum_{i} (N_{A,i} \otimes I_B) \rho_{AB} (N_{A,i} \otimes I_B)^{\dagger},
\label{eq:kraus-joint}
\end{equation}
where $\lbrace N_{A,i} \rbrace$ denotes the set of Kraus operators of the quantum channel $\mathcal{N}_A(\cdot)$ and $\lbrace N_{A,i} \otimes I_B \rbrace$ denotes the set of extended Kraus operators.

\subsection{Quantum Superdense Coding}
\label{Quantum Superdense Coding}

\begin{figure}[t]
\center
\includegraphics[width=\linewidth]{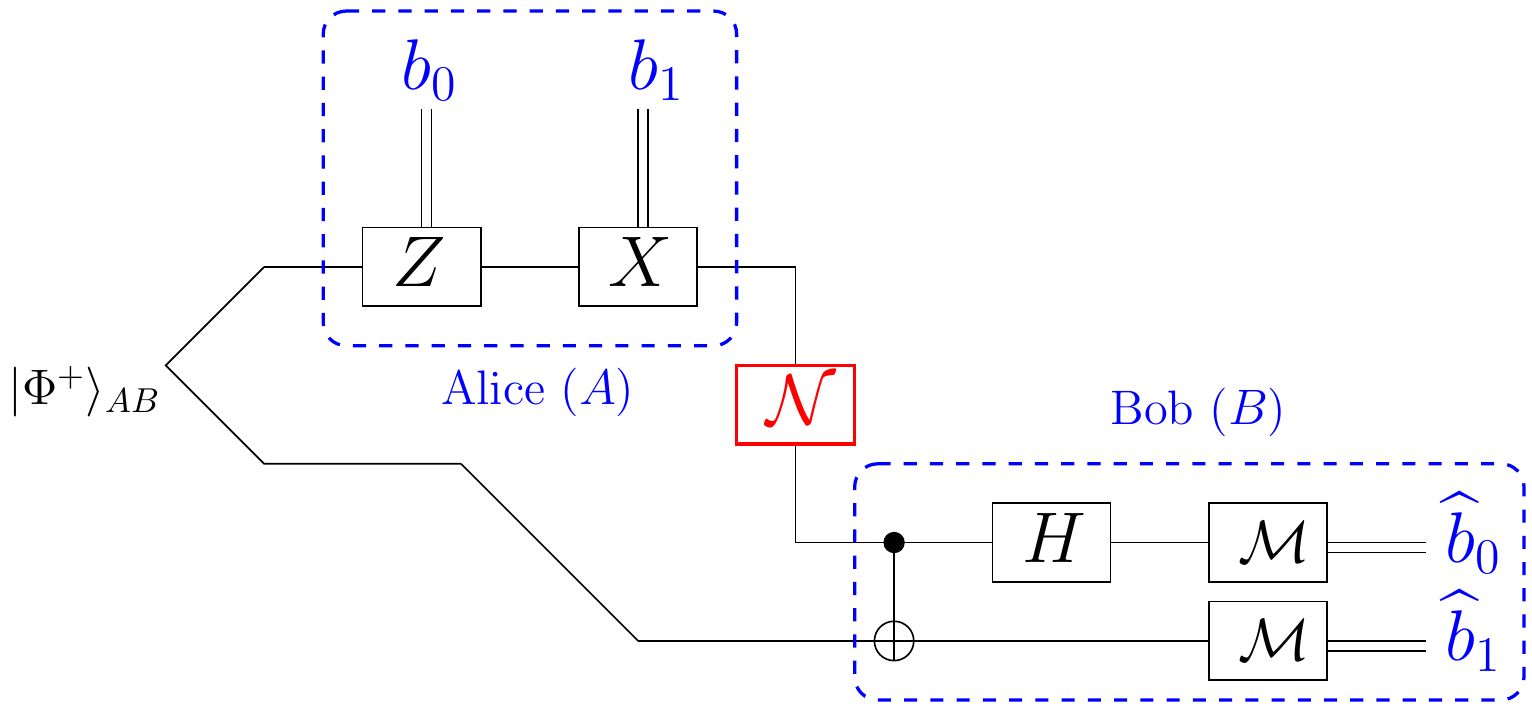}
\caption{The schematic of quantum superdense coding over noisy quantum channel $\mathcal{N}(\cdot)$. A single use of quantum channel $\mathcal{N}$ and an EPR pair $\ket{\Phi^+}_{AB}$ can be used for transferring two classical bits $\overline{b} = b_0b_1$.}
\label{fig:superdense}
\end{figure}

The main objective of this study is to evaluate the entanglement-assisted classical and quantum communication capacity over quantum trajectories. To this aim, we can determine the entanglement-assisted classical communication capacity via quantum superdense coding~\cite{bennett1992communication}, where a single use of a quantum channel and a pair of pre-shared maximally-entangled quantum state can be used for transmitting two classical bits. The general schematic of quantum superdense coding over quantum channel $\mathcal{N}(\cdot)$ is shown in Fig.~\ref{fig:superdense}. 

The quantum superdense coding protocol is commenced by pre-sharing a maximally-entangled quantum state between Alice $(A)$ and Bob $(B)$, which is constituted by the following EPR pair:
\begin{equation}
	\ket{\Phi^+}_{AB} = \frac{1}{\sqrt{2}}\left(\ket{00} + \ket{11}\right)_{AB},
\end{equation}
where the subscripts $A$ and $B$ indicate that the first qubit is held by $A$, while the second is by $B$. Let us assume that the pre-sharing step is error-free since multiple copies of EPR pairs can be prepared and hence a quantum entanglement distillation protocol can be invoked to eliminate the quantum errors~\cite{cacciapuoti2020entanglement}. 

A two-bit vector $\overline{b} = b_0b_1$ is used for applying a controlled-$Z$ and controlled-$X$ operations defined by $(Z)^{b_0}(X)^{b_1}$. To elaborate a little further, if the classical bit $b_0 = 1$, a $Z$ gate is applied to the qubit of the EPR pair on $A$ side, otherwise, an identity gate is applied. Similarly, if the classical bit $b_1 = 1$, an $X$ gate is applied, otherwise, an identity gate is applied. Consequently, the classical-to-quantum mapping of $\overline{b} = b_0b_1$ to EPR pairs is given by
\begin{align}
	b_0b_1 &= 00 \rightarrow \ket{\Phi^+}_{AB} = \frac{1}{\sqrt{2}}\left(\ket{00} + \ket{11}\right)_{AB}, \nonumber \\
	b_0b_1 &= 01 \rightarrow X_A \ket{\Phi^+}_{AB} = \frac{1}{\sqrt{2}}\left(\ket{01} + \ket{10}\right)_{AB} = \ket{\Psi^+}_{AB}, \nonumber \\
	b_0b_1 &= 10 \rightarrow Z_A \ket{\Phi^+}_{AB} = \frac{1}{\sqrt{2}}\left(\ket{00} - \ket{11}\right)_{AB} = \ket{\Phi^-}_{AB}, \nonumber \\
	b_0b_1 &= 11 \rightarrow X_A Z_A \ket{\Phi^+}_{AB} = \frac{1}{\sqrt{2}}\left(\ket{01} - \ket{10}\right)_{AB} = \ket{\Psi^-}_{AB}.
\end{align}
	
Next, the qubit of the EPR pair at $A$ side is sent through a quantum channel $\mathcal{N}(\cdot)$ as shown in Fig.~\ref{fig:superdense}. Assume that we have a quantum Pauli channel $\mathcal{N}(\cdot)$, which means that the quantum channel inflicts the bit-flip $(X)$, phase-flip $(Z)$, as well as the simultaneous bit-flip and phase-flip $(Y)$ errors with the probability of $p_x$, $p_z$, and $p_y$, respectively. Therefore, the transition probability of every possible combinations of $\overline{b} = b_0b_1$ due to $X$, $Z$, and $Y$ errors is depicted in Fig.~\ref{fig:transition}(a). Finally, as both of the qubits of the EPR pair are now at $B$ side, a Bell-state measurement is conducted to recover the two-bit vector $\widehat{b} = \widehat{b}_0\widehat{b}_1$, which is the corrupted version of $\overline{b} = b_0b_1$. With this, the quantum superdense coding protocol over quantum channel $\mathcal{N}(\cdot)$ is completed.

\subsection{The Capacity of Entanglement-Assisted Classical Communication}
\label{The Capacity of Entanglement-Assisted Classical Communication}

Evaluating the capacity of quantum superdense coding protocol can be reformulated as evaluating the capacity of quaternary discrete classical communication. To obtain a single-letter formula of its capacity, first, we provide the general definition of classical communication capacity.

\begin{definition} \textit{Classical communication capacity}~\cite{shannon1948mathematical}. The classical communication capacity is defined by 
\begin{equation}
C = \max_{P(x)} I(X;Y) = \max_{P(x)} \lbrace H(Y) - H(Y|X) \rbrace,
\label{eq:capacity-general}
\end{equation}
where $I(X;Y)$ is the mutual information between random variables $X$ and $Y$, $P(x)$ is the probability distribution of the source emitting symbol $x \in X$, $H(Y)$ is the entropy of random variable $Y$, $H(Y|X)$ is the conditional entropy of random variable $Y$ conditioned by $X$.
\end{definition}

Furthermore, the capacity of~\eqref{eq:capacity-general} is also equivalent to 
\begin{align}
C &= \log_2 M - \sum_{i,j} \displaystyle P(x_j) H(Y|X = x_j) \nonumber \\
 &= \log_2 M - \sum_{j} \displaystyle P(x_j) \sum_{i} P(y_i|x_j) \log_2 \frac{1}{\displaystyle P(y_i|x_j)}.
\label{eq:capacity-specific}
\end{align}
Since we assume that we have an equiprobable source for the symbols $x \in X$ and $P(y_i|x_j)$ is time invariant, the expression of~\eqref{eq:capacity-specific} can be further simplified to
\begin{equation}
C = \log_2 M + \sum_{i} P(y_i|x) \log_2 P(y_i|x),
\label{eq:capacity-specific-simple}
\end{equation}
where $P(y_i|x)$ is the transition probability of $y_i$ for any $x \in X$ characterized by the channel\footnote{The transition probability of $P(y_i|x)$ for quantum Pauli channel is known to be symmetric for any $x \in X$.}.

\begin{figure}[t]
\center
\includegraphics[width=\linewidth]{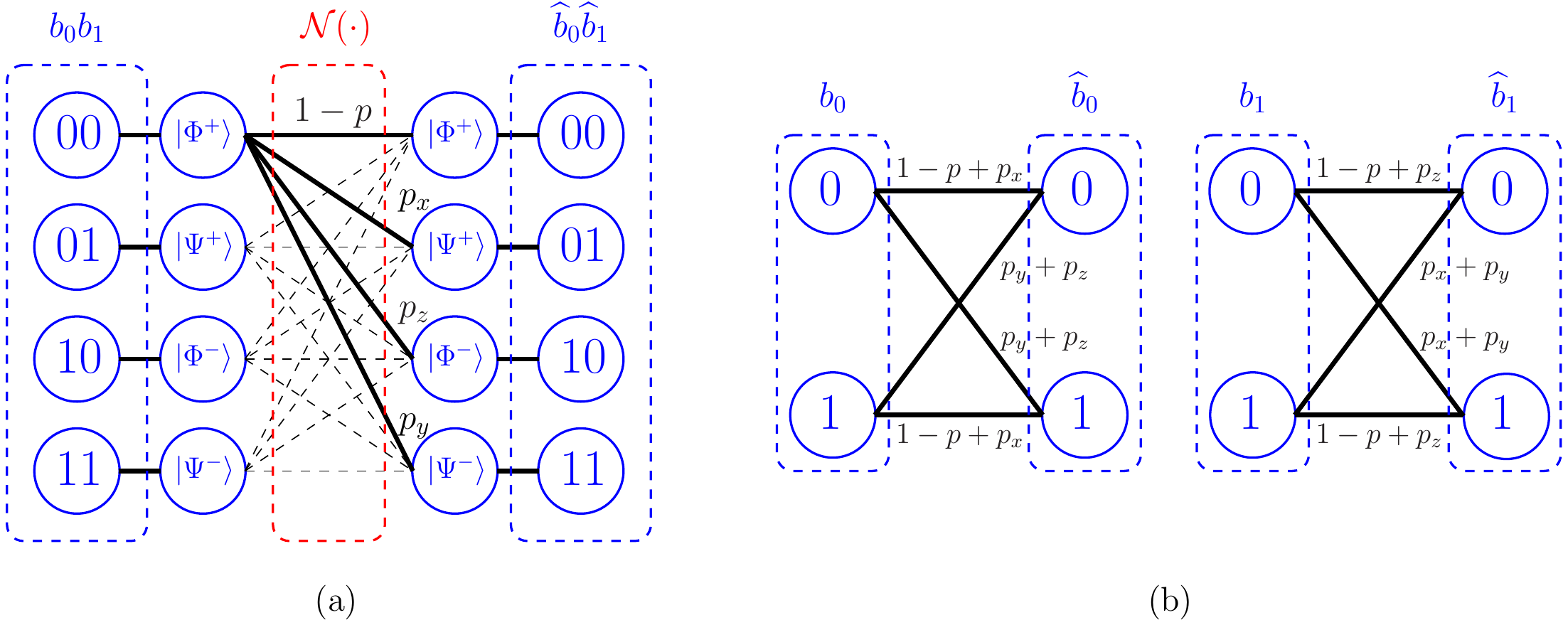}
\caption{The transition probability of $\overline{b} = b_0b_1$ over quantum Pauli channels $\mathcal{N}(\cdot)$, which will be utilized for deriving the entanglement-assisted (a) symbol-based and (b) bit-based classical communication capacity. Compared to (a), where the transition probability is derived jointly for $\overline{b} = b_0b_1$, each of the transition probabilities in (b) is derived for $b_0$ and $b_1$.}
\label{fig:transition}
\end{figure}

In order to glean a clearer idea, let us proceed with an example. Consider the quantum depolarizing channel $\mathcal{N}(\cdot)$ affecting a single-qubit having the density matrix $\rho$ as follows:
\begin{equation}
\mathcal{N}(\rho) = (1 - p)\rho + \frac{p}{3} \left( {X}\rho{X} + {Y}\rho{Y} + {Z}\rho{Z} \right),
\label{eq:quantum-depolarizing}
\end{equation}
where $p$ is the depolarizing probability. Now, we have to make the connection between the single-qubit depolarizing channel of~\eqref{eq:quantum-depolarizing} and the transition probability of Fig.~\ref{fig:transition}. Utilizing the extended mapping of the quantum channel of~\eqref{eq:kraus-joint} and by considering $\rho_{AB} = \ket{\Phi^+}_{AB} \bra{\Phi^+}_{AB}$, the effect of a single-qubit depolarizing channel on an EPR pair $\rho_{AB}$ is given by 
\begin{align}
(\mathcal{N}_{A} \otimes I_{B})(\rho_{AB}) &= \left( 1 - p \right)\ket{\Phi^+} \bra{\Phi^+} + \frac{p}{3} \ket{\Psi^+} \bra{\Psi^+} \nonumber \\
 &+ \frac{p}{3} \ket{\Psi^-} \bra{\Psi^-} + \frac{p}{3} \ket{\Phi^-} \bra{\Phi^-}.
\label{eq:transition-epr}
\end{align}
Based on~\eqref{eq:transition-epr}, we obtain the transition probability $p(y_i|x)$ as follows:
\begin{equation}
p(y_i|x) = \left\{
\begin{array}{ll}
1 - p, & \text{for} \ i = 0 \\
p_x = p_y = p_z = \dfrac{p}{3}, & \text{for} \ i = 1,2,3
\end{array},
\right.
\label{eq:transition-epr-binary}
\end{equation}
where $i$ represents the decimal representation of binary vector $\overline{b} = b_0b_1$. Finally, substituting the value $M = 4$ and $p(y_i|x)$ of~\eqref{eq:transition-epr-binary} into~\eqref{eq:capacity-specific-simple}, we obtain the entanglement-assisted classical communication capacity in terms of $p$ as follows~\cite{bennett1999entanglement}:
\begin{equation}
C_{E,s} = 2 + (1-p) \log_2 (1 - p) + p \log_2 \frac{p}{3}.
\label{eq:symbol-based}
\end{equation}
For a single-qubit depolarizing channel, the capacity can be derived for both symbol-based and bit-based classical communication. The symbol-based capacity is already given in~\eqref{eq:symbol-based} based on the transition probability of $b_0b_1$ in Fig.~\ref{fig:transition}(a). By contrast, the bit-based capacity can be determined by summing the individual capacity of $b_0$ and $b_1$ based on the transition probability of Fig.~\ref{fig:transition}(b), where we obtain
\begin{equation}
C_{E,b} = 2 + \left( 2-\dfrac{4p}{3} \right) \log_2 \left( 1 - \frac{2p}{3} \right) + \dfrac{4p}{3} \log_2 \frac{2p}{3}.
\label{eq:bit-based}
\end{equation}
As a benchmark, we provide the unassisted classical communication capacity of a single-qubit depolarizing channel. Luckily, for a single-qubit depolarizing channel, the capacity always correspond to the Von Neumann measurements~\cite{shor2003capacities}, where in terms of $p$ is given by
\begin{equation}
C = 1 + \left( 1-\dfrac{2p}{3} \right) \log_2 \left( 1 - \frac{2p}{3} \right) + \dfrac{2p}{3} \log_2 \frac{2p}{3}.
\label{eq:unassisted}
\end{equation}
Notice that the unassisted classical capacity is exactly half of the bit-based entanglement-assisted classical capacity\footnote{With reference to the entanglement-assisted capacity, this statement is true since the cost of distributing entanglement is not taken into account, as generally assumed in the theoretical quantum information theory framework.}.

\begin{remark} The inherent inter-bit correlation within the symbols makes the symbol-based capacity slightly higher than the sum of individual bit-based capacity. Consequently, the entanglement-assisted classical communication capacity is also slightly higher than the sum of individual unassisted classical communication capacity.
\end{remark}

\begin{figure}[t!]
\center
\includegraphics[width=\linewidth]{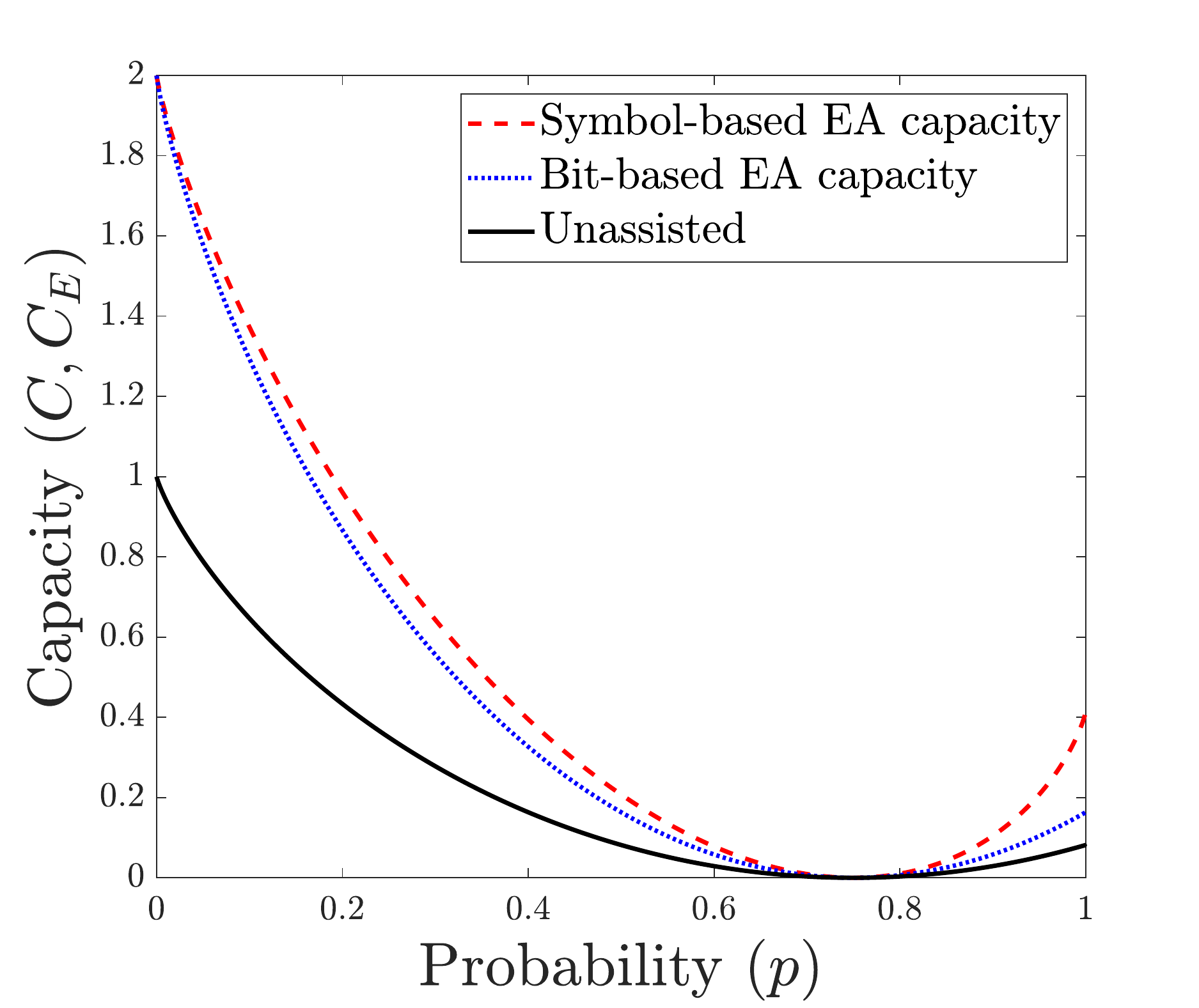}
\caption{The entanglement-assisted classical communication capacity over a single use of quantum depolarizing channel. The plots are based on~\eqref{eq:symbol-based},~\eqref{eq:bit-based}, and~\eqref{eq:unassisted}. EA stands for entanglement-assisted.}
\label{fig:capacity-single-depolarizing}
\end{figure}

We portray the symbol-based capacity of~\eqref{eq:symbol-based}, the bit-based capacity of~\eqref{eq:bit-based}, and the unassisted capacity of~\eqref{eq:unassisted} in Fig.~\ref{fig:capacity-single-depolarizing}. Observe that all the capacities are equal to $0$ when $p = 0.75$. This specific point is associated with the fully-depolarizing quantum channel where we have $\mathcal{N}(\rho) = I/2$ for any arbitrary $\rho$. Consequently, there is no classical information can be transmitted, either by employing entanglement-assisted or unassisted, through a fully-depolarizing quantum channel.

Finally, following the reasoning we utilized for deriving the entanglement-assisted classical communication capacity for single-qubit depolarizing channel, we may also directly derive the entanglement-assisted classical communication capacity for the general quantum Pauli channel of~\eqref{eq:quantum‐pauli‐channel}, which is given by 
\begin{equation}
	C_E = 2 + p_0 \log_2 p_0 + p_1 \log_2 p_1 + p_2 \log_2 p_2 + p_3 \log_2 p_3.
	\label{eq:capacity-pauli}
\end{equation}
For the rest of treatise, we are going to use the entanglement-assisted capacity of~\eqref{eq:capacity-pauli} for evaluating the bottleneck capacity of the quantum channels arranged in a well-defined causal order, which we will elaborate in the next section.

\section{The Classical and Quantum Trajectory}
\label{The Classical and Quantum Trajectory}

As we have briefly alluded in Section~\ref{Introduction}, the unique properties of quantum information allow the information carrier to traverse multiple classical trajectories simultaneously. In this section, we provide the formal mathematical description of the classical and quantum trajectory of two quantum channels. More specifically, a classical trajectory of two quantum channels is characterized by a well-defined causal order of either $\mathcal{D}(\cdot) \rightarrow \mathcal{E}(\cdot)$ or $\mathcal{E}(\cdot) \rightarrow \mathcal{D}(\cdot)$. By contrast, a quantum trajectory of two quantum channels is characterized by the superposition of two classical trajectories $\mathcal{D}(\cdot) \rightarrow \mathcal{E}(\cdot)$ and $\mathcal{E}(\cdot) \rightarrow \mathcal{D}(\cdot)$ implying that the trajectory exhibits an indefinite causal order of the quantum channels $\mathcal{D}(\cdot)$ and $\mathcal{E}(\cdot)$.

\subsection{Classical Trajectory}
\label{Classical Trajectory}

Consider two quantum channels $\mathcal{D}(\cdot)$ and $\mathcal{E}(\cdot)$ as follows:
\begin{equation}
	\mathcal{D}(\rho) = \sum_{i} D_i \rho D_i^{\dagger}, \quad \mathcal{E}(\rho) = \sum_{j} E_j \rho E_j^{\dagger}.
\end{equation}
The resultant channel $\mathcal{S}(\cdot)$ of two quantum channels over a classical trajectory with a definite causal order of $\mathcal{D}(\cdot) \rightarrow \mathcal{E}(\cdot)$ is formulated as
\begin{equation}
	\mathcal{S}(\mathcal{D},\mathcal{E})(\rho) = \sum_{i,j} W_{i,j} \rho W_{i,j}^{\dagger},
\end{equation}
where the Kraus operators are given by
\begin{equation}
	W_{i,j} = E_j D_i.
	\label{eq:kraus-serial}
\end{equation}

The amount of transferable information over a classical trajectory of two quantum channels $\mathcal{D}(\cdot)$ and $\mathcal{E}(\cdot)$ is upper-bounded by the so-called bottleneck capacity, which applies for both classical and quantum communications as well as for both unassisted and entanglement-assisted communications.

\begin{definition}\textit{The bottleneck capacity}~\cite{wilde2013entanglement, cacciapuoti2019capacity}. The communication capacity over classical trajectory of two quantum channels is upper-bounded by
	\begin{equation}
		C_{\text{B}} = \min \lbrace C(\mathcal{D}), C(\mathcal{E}) \rbrace,
		\label{eq:bottleneck}
	\end{equation} 
where $C(\mathcal{D})$ and $C(\mathcal{E})$ are the capacities of the individual quantum channels $\mathcal{D}(\cdot)$ and $\mathcal{E}(\cdot)$, respectively. When $C(\mathcal{D})$ and $C(\mathcal{E})$ represent the entanglement-assisted communication capacities of $\mathcal{D}(\cdot)$ and $\mathcal{E}(\cdot)$, respectively, the notation $C_{\text{E,B}}$ is used for portraying the associated bottleneck capacity.
\label{def:bottleneck}
\end{definition}

In this treatise, we consider $\mathcal{D}(\cdot)$ and $\mathcal{E}(\cdot)$ to be general quantum Pauli channels of~\eqref{eq:quantum‐pauli‐channel}. Therefore, we may evaluate the individual $C(\mathcal{D})$ and $C(\mathcal{E})$ using the entanglement-assisted classical communication capacity for quantum Pauli channels of~\eqref{eq:capacity-pauli} to obtain the bottleneck capacity $C_{\text{E,B}}$.

\subsection{Quantum Trajectory}
\label{Quantum Trajectory}

Given two quantum channels $\mathcal{D}(\cdot)$ and $\mathcal{E}(\cdot)$, the resultant quantum channel $\mathcal{S}(\cdot)$ over quantum trajectory is formulated as~\cite{salek2018quantum}
\begin{equation}
	\mathcal{S}_{\omega} (\mathcal{D},\mathcal{E})(\rho) = \sum_{i,j} W_{i,j} (\rho \otimes \omega) W_{i,j}^{\dagger},
	\label{eq:resultant}
\end{equation}
where $\omega$ is the control qubit and the Kraus operators are given by 
\begin{equation}
	W_{i,j} = E_j D_i \otimes \ket{0} \bra{0} + D_i E_j \otimes \ket{1} \bra{1}.
	\label{eq:kraus-time-superposition}
\end{equation}
The formulation of the Kraus operators over quantum trajectory implies that we may create a superposition of classical trajectories using the control qubit $\omega$. More specifically, when $\omega = \ket{0} \bra{0}$, the information carrier of the quantum state $\rho$ traverses the classical trajectory of $\mathcal{D}(\cdot) \rightarrow \mathcal{E}(\cdot)$. By contrast, when $\omega = \ket{1} \bra{1}$, the information carrier of the quantum state $\rho$ traverses the classical trajectory of $\mathcal{E}(\cdot) \rightarrow \mathcal{D}(\cdot)$. For instance, we may create an equal superposition of both classical trajectories by initializing $\omega = \ket{+} \bra{+}$, which means that the quantum state $\rho$ is traversed through both classical trajectories simultaneously, which results in an indefinite causal order of quantum channels. Thus, throughout this treatise, we assume that the control qubit is always initialized in the quantum state of $\omega = \ket{+} \bra{+}$ since it gives us the capability of detecting the superposition of causal orders from anti-commuting Kraus operators and ultimately provides us with the highest possible capacity gain~\cite{salek2018quantum}.

More specifically, let us assume that $D_1$ and $E_1$ are two anti-commuting Kraus operators, i.e. $D_1 E_1 = -E_1 D_1$. Thus, the Kraus operators of Eq~\eqref{eq:kraus-time-superposition} can be expressed as follows:
\begin{align}
	W_{1,1} &= E_1 D_1 \otimes \ket{0} \bra{0} - E_1 D_1 \otimes \ket{1} \bra{1} \nonumber \\
 	&= E_1 D_1 \otimes (\ket{0} \bra{0} - \ket{1} \bra{1}) = E_1 D_1 \otimes Z.
	\label{eq:kraus-time-superposition-example}
\end{align}
Therefore, the action of Kraus operators $W_{1,1}$ of~\eqref{eq:kraus-time-superposition-example} on the initial quantum state of $\rho \otimes \ket{+} \bra{+}$ is given by
\begin{align}
	W_{1,1} \left(\rho \otimes \ket{+} \bra{+}\right) W_{1,1}^{\dagger} &= (E_1 D_1)\rho(E_1 D_1)^{\dagger} \otimes \left( Z \ket{+} \bra{+} Z \right) \nonumber \\
 	&= (E_1 D_1)\rho(E_1 D_1)^{\dagger} \otimes \ket{-} \bra{-}.
\end{align}
According to this result, when we measure the quantum state $\ket{-} \bra{-}$ on the control qubit, we can infer that a superposition of causal orders from two anti-commuting Kraus operators has taken place. Thus, we can utilize the measurement result to our advantage for improving the performance of classical and quantum communication.

\begin{remark}
	If the control qubit is initialized in the quantum state of $\omega = \ket{+} \bra{+}$, the superposition of causal orders from two anti-commuting Kraus operators transforms the control qubit into $\ket{-} \bra{-}$.
\end{remark}

Based on the formal description of the quantum trajectory in~\eqref{eq:resultant}, indeed we require an additional auxiliary qubit $\omega$ to control the superposition of the causal orders of the quantum channels $\mathcal{D}(\cdot)$ and $\mathcal{E}(\cdot)$. Intuitively, we have the inclination to make a comparison between the advocated scheme presented in this treatise to another auxiliary-qubits assisted scheme, such as quantum error-correction codes~\cite{babar2018duality, chandra2020direct}. Nevertheless, viewing the control qubit $\omega$ of~\eqref{eq:resultant} in the same light as the auxiliary qubits in quantum error-correction codes can be very problematic. More specifically, the auxiliary qubits in quantum error-correction codes are encoded together with the logical qubits and sent through the quantum channels carrying a certain amount of information. Consequently, the incorporation of auxiliary qubits in quantum error-correction codes increases the total number of quantum channel uses. By contrast, the utilization of the auxiliary qubit $\omega$ of~\eqref{eq:resultant} does not increase the number of quantum channel uses. In fact, as shown in~\cite{kristjansson2020resource}, the encoding operation of~\eqref{eq:resultant} must be considered as a \textit{non-side-channel generating} operation, since the control qubit $\omega$ does not carry -- or embed in any way -- the information from the source to the destination. The initial state of the control qubit $\omega$ is fixed as part of the placement and thus, it is deemed inaccessible to the sender for encoding information. Having said that, we refer enthusiastic readers to~\cite{kristjansson2020resource} for the full discourse on quantum resource theories.

\section{The Entanglement-Assisted Capacity over Classical and Quantum Trajectory}
\label{The Entanglement-Assisted Capacity over Classical and Quantum Trajectory}

We employ the formulation of classical and quantum trajectories of~\eqref{eq:kraus-serial} and~\eqref{eq:kraus-time-superposition} and incorporate them into~\eqref{eq:kraus-joint} for devising the transition probability required for calculating the classical communication capacity of~\eqref{eq:capacity-specific-simple}. Consider the following quantum Pauli channels $\mathcal{D}(\cdot)$ and $\mathcal{E}(\cdot)$:
\begin{equation}
	\mathcal{D}(\rho) = p_0 \rho + p_1 X\rho X + p_2 Y\rho Y + p_3 Z\rho Z, \label{eq:channeld}
\end{equation}
\begin{equation}
	\mathcal{E}(\rho) = q_0 \rho + q_1 X\rho X + q_2 Y\rho Y + q_3 Z\rho Z. \label{eq:channele}
\end{equation}
Our results of the entanglement-assisted classical communication capacity over classical and quantum trajectories are summarized in the following propositions.

\begin{proposition} 
	The entanglement-assisted classical communication capacity $C_{\text{E,C}}$ of the two arbitrary quantum Pauli channels in~\eqref{eq:channeld} and~\eqref{eq:channele} over a classical trajectory is given by
	\begin{equation}
		C_{\text{E,C}} = 2 + {A}_0 \log_2 {A}_0 + {A}_1 \log_2 {A}_1 + {A}_2 \log_2 {A}_2 + {A}_3 \log_2 {A}_3,
		\label{eq:capacity-serial-general}
	\end{equation}
where ${A}_0 = p_0 q_0 + p_1 q_1 + p_2 q_2 + p_3 q_3$, ${A}_1 = p_0 q_1 + p_1 q_0 + p_2 q_3 + p_3 q_2$, ${A}_2 = p_0 q_2 + p_2 q_0 + p_3 q_1 + p_1 q_3$, and ${A}_3 = p_0 q_3 + p_3 q_0 + p_1 q_2 + p_2 q_1$.
	\begin{IEEEproof}
		Please refer to Appendix~\ref{Proof of Proposition Serial}.
	\end{IEEEproof}
	\label{prop_1}
\end{proposition}

\begin{proposition}
	The entanglement-assisted classical communication capacity $C_{\text{E,Q}}$ of the two arbitrary quantum Pauli channels of~\eqref{eq:channeld} and~\eqref{eq:channele} over a quantum trajectory is given by
	\begin{align}
		C_{\text{E,Q}} &= 2 + H(\alpha) + {A}_0 \log_2 {A}_0 \nonumber \\
		&+ {A}_1^+ \log_2 {A}_1^+ + {A}_2^+ \log_2 {A}_2^+ + {A}_3^+ \log_2 {A}_3^+ \nonumber \\
		&+ {A}_1^- \log_2 {A}_1^- + {A}_2^- \log_2 {A}_2^- + {A}_3^- \log_2 {A}_3^-,
		\label{eq:capacity-switch-general}
	\end{align}
where $\alpha = p_{\ket{+}} = {A}_0 + {A}_1^+ + {A}_2^+ + {A}_3^+$, ${A}_0 = p_0 q_0 + p_1 q_1 + p_2 q_2 + p_3 q_3$, ${A}_1^+ = p_0 q_1 + p_1 q_0$, ${A}_2^+ = p_0 q_2 + p_2 q_0$, ${A}_3^+ = p_0 q_3 + p_3 q_0 $, ${A}_1^- = p_2 q_3 + p_3 q_2$, ${A}_2^- = p_3 q_1 + p_1 q_3$, ${A}_3^- = p_1 q_2 + p_2 q_1$, and $H(\alpha)$ is the binary entropy of $\alpha$ defined by $- \alpha \log_2 \alpha - (1 - \alpha) \log_2 (1 - \alpha)$.
	\begin{IEEEproof}
		Please refer to Appendix~\ref{Proof of Proposition Switch}.
	\end{IEEEproof}
	\label{prop_2}
\end{proposition}

Observe from~\eqref{eq:capacity-serial-general} and~\eqref{eq:capacity-switch-general}, the following inequality always holds: $C_{\text{E,Q}} \geq C_{\text{E,C}}$. The equality holds when ${A}_1^- = {A}_2^- = {A}_3^- = 0$ implying ${A}_0^+ = {A}_0$, ${A}_1^+ = {A}_1$, ${A}_2^+ = {A}_2$, ${A}_3^+ = {A}_3$, and $H(\alpha) = 0$. It means that the advantage of quantum trajectory can only be observed when the coefficient pair of $\{ p_i, q_j | i,j = 1, 2, 3\}$ for non-commuting Pauli matrices is non zero. More specifically, the two classical trajectories of two non-commuting Pauli matrices, for example $XZ = -ZX$, transforms the control qubit $\omega = \ket{+} \bra{+}$ into $\omega = \ket{-} \bra{-}$, which consequently increases the entanglement-assisted classical communication capacity compared to that of classical trajectories.

\begin{remark}
In order to glean the advantage of quantum trajectory for increasing the entanglement-assisted classical communication capacity of two arbitrary quantum Pauli channels $\mathcal{D}(\cdot)$ and $\mathcal{E}(\cdot)$, the coefficient pair of $\{ p_i, q_j | i,j = 1, 2, 3\}$ for non-commuting Pauli matrices has to be non-zero.
\end{remark}

\subsection{Classical Communication Capacity}
\label{Classical Communication Capacity}

In the following subsections, we provide several derivative results of Proposition~\ref{prop_1} and~\ref{prop_2} for various types of quantum Pauli channels, which are widely considered in the practical applications of quantum communications. More specifically, we consider the combination of bit-flip and phase flip quantum channels, quantum entanglement-breaking channels, as well as quantum depolarizing channels. Moreover, the experimental implementation of the indefinite causal orders of the aforementioned quantum channels can be found in~\cite{guo2020experimental, goswami2020experiments, goswami2020increasing}.

\subsubsection{Quantum Bit-Flip and Phase-Flip Channels}
\label{Bit-Flip and Phase-Flip Channels}

We consider two quantum Pauli channels constituted by bit-flip and phase-flip quantum channels. A quantum bit-flip channel is defined as follows:
\begin{equation}
\mathcal{D}(\rho) = (1-p)\rho + p X \rho X,
\label{eq:kraus-bit-flip}
\end{equation}
where the Kraus operators are given by $D_1 = \sqrt{1-p}I$ and $D_2 = \sqrt{p}X$. A quantum phase-flip channel is defined as follows:
\begin{equation}
\mathcal{E}(\rho) = (1-q)\rho + q Z \rho Z,
\label{eq:kraus-phase-flip}
\end{equation}
where the Kraus operators are given by $E_1 = \sqrt{1-q}I$ and $E_2 = \sqrt{q}Z$. Based on the Kraus operators of the bit-flip and phase-flip quantum channels of~\eqref{eq:kraus-bit-flip} and~\eqref{eq:kraus-phase-flip}, respectively, and based on the formulation of Kraus operators for classical trajectory of~\eqref{eq:kraus-serial}, we obtain the following corollary.
\begin{corollary}
\label{col_1}
The entanglement-assisted classical communication capacity $C_{\text{E,C}}$ of bit-flip and phase-flip quantum channels over classical trajectory is given by
\begin{align}
C_{\text{E,C}} &= 2 + \left[(1-p)(1-q)\right] \log_2 \left[(1-p)(1-q)\right] \nonumber \\ 
&+ \left[p(1-q)\right] \log_2 \left[p(1-q)\right] + \left[(1-p)q\right] \log_2 \left[(1-p)q\right] \nonumber \\
&+ pq \log_2 pq.
\label{eq:capacity-serial-bit-phase}
\end{align}
\begin{IEEEproof}
By substituting $p_0 = 1 - p$, $q_0 = 1 - q$, $p_1 = p$, $q_3 = q$, and the rest of the coefficients equal to $0$ into~\eqref{eq:capacity-serial-general} of Proposition~\ref{prop_1}, we obtain the result in~\eqref{eq:capacity-serial-bit-phase}.
\end{IEEEproof}
\end{corollary}
Similarly, based on the formulation of Kraus operators for quantum trajectory of~\eqref{eq:kraus-time-superposition}, we also obtain the following corollary.
\begin{corollary}
\label{col_2}
The entanglement-assisted classical communication capacity $C_{\text{E,Q}}$ of bit-flip and phase-flip quantum channels over quantum trajectory is given by
\begin{align}
C_{\text{E,Q}} &= 2 + H(\alpha) + \left[(1-p)(1-q)\right] \log_2 \left[(1-p)(1-q)\right] \nonumber \\
&+ \left[p(1-q)\right] \log_2 \left[p(1-q)\right] + \left[(1-p)q\right] \log_2 \left[(1-p)q\right] \nonumber \\ 
&+ pq \log_2 pq,
\label{eq:capacity-switch-bit-phase}
\end{align}
where $\alpha = p_{\ket{+}}= 1 - pq$.
\begin{IEEEproof}
By substituting $p_0 = 1 - p$, $q_0 = 1 - q$, $p_1 = p$, $q_3 = q$, and the rest of the coefficients equal to $0$ into~\eqref{eq:capacity-switch-general} of Proposition~\ref{prop_2}, we obtain the result in~\eqref{eq:capacity-switch-bit-phase}. 
\end{IEEEproof}
\end{corollary}

\begin{figure}[t]
\center
\includegraphics[width=\linewidth]{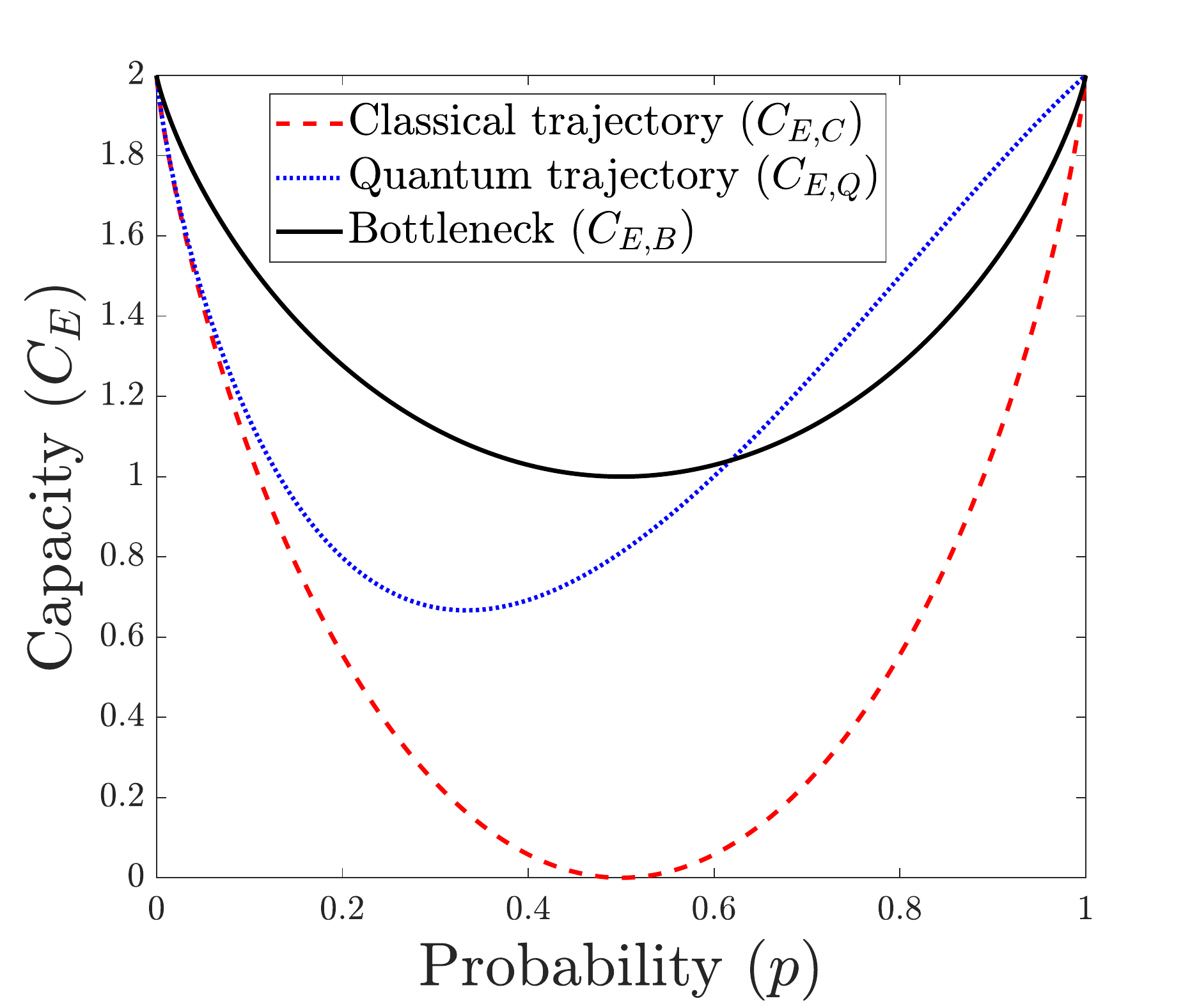}
\caption{The entanglement-assisted classical communication capacity for the combination of bit-flip and phase-flip quantum channels. These plots are based on~\eqref{eq:capacity-serial-bit-phase} and~\eqref{eq:capacity-switch-bit-phase}.}
\label{fig:capacity-bit-phase-flip}
\end{figure}

To demonstrate the advantage of the quantum trajectory compared to the classical trajectory, let us consider a scenario where we have $p = q$. We plot the entanglement-assisted capacity of~\eqref{eq:capacity-serial-bit-phase} and~\eqref{eq:capacity-switch-bit-phase} in a scenario of $p = q$ in Fig.~\ref{fig:capacity-bit-phase-flip}. In addition to these results, we include the bottleneck capacity of the two quantum channels. Observe that the capacity of the resultant channel over classical trajectory indeed cannot violate the stringent bottleneck capacity. However, the capacity over quantum trajectory violates the bottleneck capacity in the region of $0.618 < p < 1$. This violation of bottleneck capacity is induced by the combination of two causal orders of two non-commuting Pauli matrices from the Kraus operators of the quantum channels, $XZ = -ZX$. This combination transforms the control qubit $\omega$, which is initialized in the quantum state of $\ket{+} \bra{+}$, into quantum state of $\ket{-} \bra{-}$. Consequently, this gives us a perfect entanglement-assisted classical communication with a probability of $p^2$ since every time we measure $\ket{-} \bra{-}$ on the control qubit, we can apply $XZ$ operation to perfectly recover the quantum state of information qubit $\rho$. This is very important to highlight that the measurement of the control qubit $\omega$ does not give us any information about which causal order the information carrier has traversed, or in other words it means the measurement does not collapse the superposition of the classical trajectories. Instead, what the measurement tells us is if the quantum channels have inflicted two non-commuting Pauli matrices in two different causal orders, which can only occur when the information carrier has actually traversed both classical trajectories simultaneously.

\begin{remark}
The result in Corollary~\ref{col_2} implies that the indefinite causal order of the quantum channels allows us to violate the bottleneck capacity, which constrains the capacity of classical and quantum communication traversed over a classical trajectory with a definite causal order of quantum channels.
\end{remark} 

For the sake of demonstration, the results presented in in Fig.~\ref{fig:capacity-bit-phase-flip} is valid for a scenario of $p = q$. However, the results presented in Corollary~\ref{col_1} and~\ref{col_2} are actually general for a wider range of $p$ and $q$. To enrich our discussion and to explore the advantages gleaned by the indefinite causal order of quantum channels, we introduce two additional metrics, namely the capacity gain and the bottleneck violation based on the following definitions.

\begin{definition}
The capacity gain $G$ is defined as the difference between the entanglement-assisted classical communication capacity over quantum trajectory and that of classical trajectory, which depicts the amount of additional classical information that can be sent by exploiting the indefinite causal order of quantum channels. Formally, the capacity gain $G$ can be expressed as
\begin{equation}
G = C_{\text{E,Q}} - C_{\text{E,C}},
\end{equation} 
where $C_{\text{E,Q}}$ is the entanglement-assisted classical communication capacity over quantum trajectory presented in~\eqref{eq:capacity-switch-general} of Proposition~\ref{prop_2}, while $C_{\text{E,C}}$ is that of over classical trajectory obtained from~\eqref{eq:capacity-serial-general} of Proposition~\ref{prop_1}.
\end{definition}

\begin{definition}
The bottleneck violation $V$ is defined as the non-negative gain obtained by entanglement-assisted classical communication capacity over quantum trajectory against the bottleneck capacity. The bottleneck violation $V$ depicts the amount of capacity gain which cannot be attained through any definite causal order of quantum channels, which is formally expressed as
\begin{equation}
V = \max \lbrace 0, C_{\text{E,Q}} - C_{\text{E,B}} \rbrace,
\end{equation} 
where $C_{\text{E,Q}}$ is the entanglement-assisted classical communication capacity over quantum trajectory presented in~\eqref{eq:capacity-switch-general} of Proposition~\ref{prop_2}, while $C_{\text{E,B}}$ is the bottleneck capacity of entanglement-assisted classical communication based on~\eqref{eq:bottleneck} of Definition~\ref{def:bottleneck}.
\end{definition}

\begin{figure*}[t]
    \centering
    \begin{subfigure}{0.49\linewidth}
        \includegraphics[width=\linewidth]{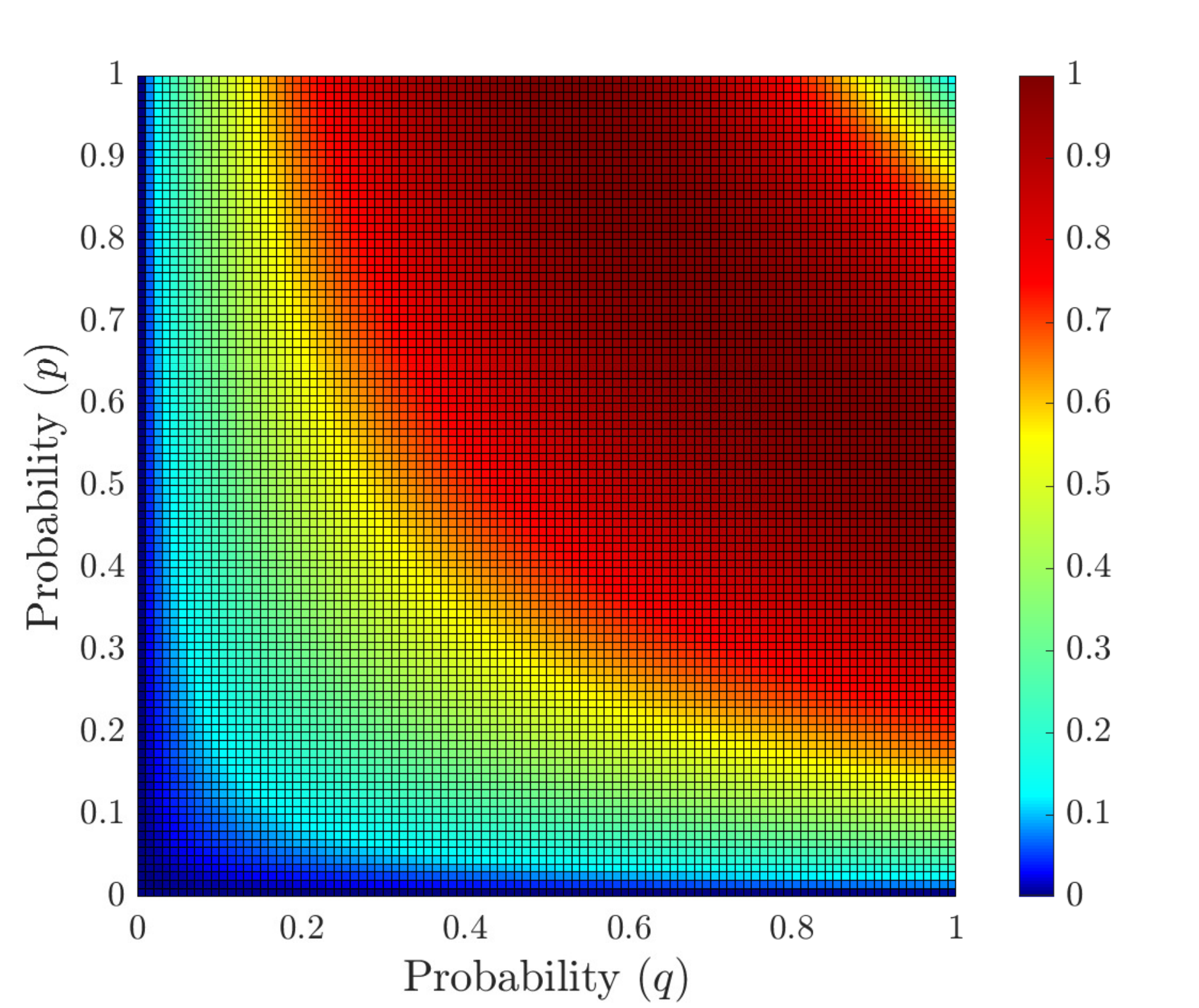}
		\caption{Capacity gain ($G$)}
		\label{fig:capacity-switch-bit-phase-flip-gain}
    \end{subfigure}
    \hfill
    \begin{subfigure}{0.49\linewidth}
        \includegraphics[width=\linewidth]{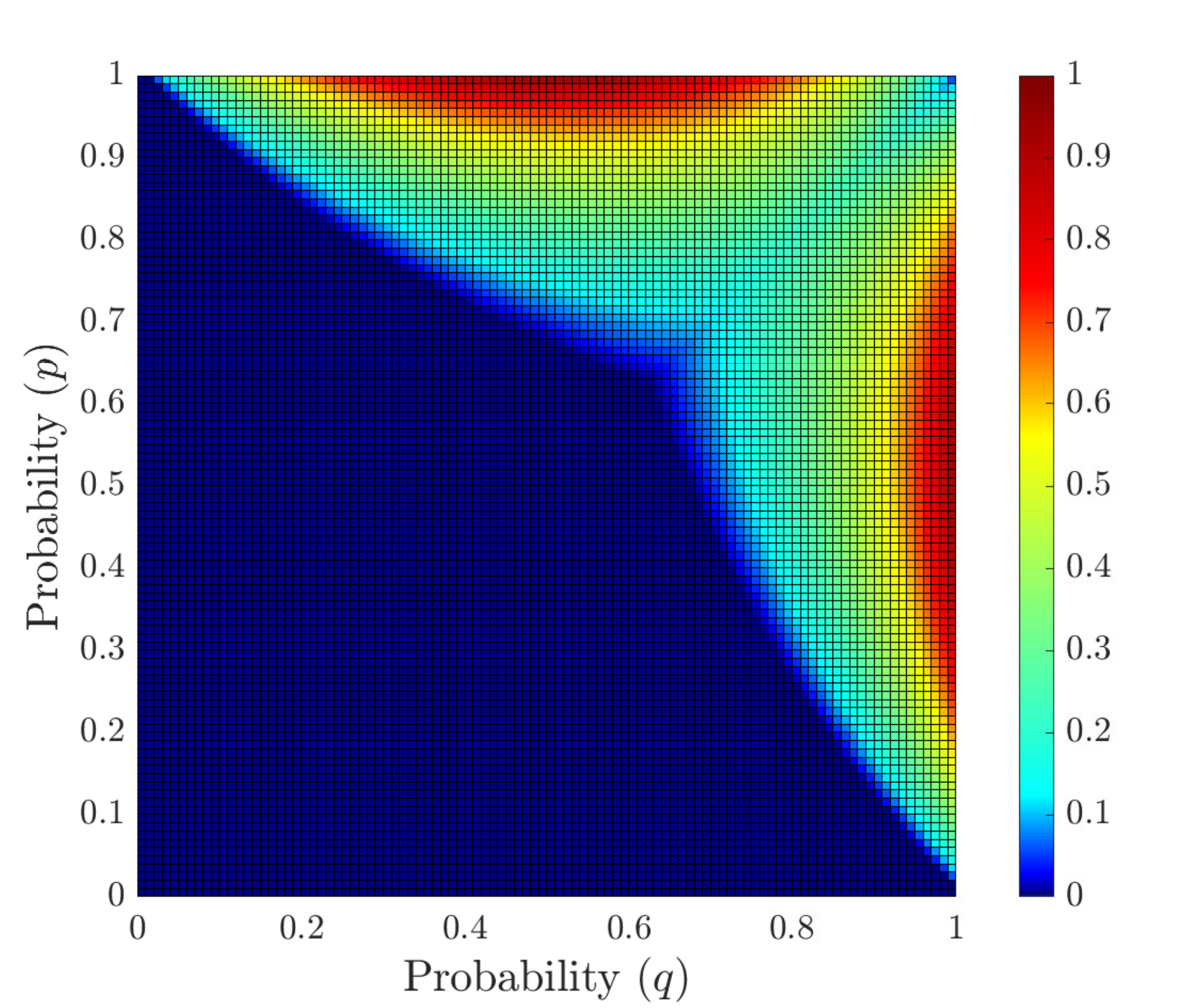}
		\caption{Bottleneck violation ($V$)}
		\label{fig:capacity-switch-bit-phase-flip-violation}
    \end{subfigure}
\caption{(a) The entanglement-assisted classical capacity gain of the bit-flip and phase-flip quantum channels over quantum trajectory against classical trajectory. (b) The violation of bottleneck capacity due to the indefinite causal order of quantum channels.}
\label{fig:capacity-bit-phase-flip-pq}
\hrulefill
\end{figure*}

We portray the capacity gain for the bit-flip and phase-flip quantum channels over quantum trajectory in Fig.~\ref{fig:capacity-bit-phase-flip-pq}(a), while the bottleneck violation in Fig.~\ref{fig:capacity-bit-phase-flip-pq}(b). Observe that based on the color map in Fig.~\ref{fig:capacity-bit-phase-flip-pq}(a), we obtain the benefit of capacity gain at all range values of $0 < p,q < 1$. However, the significant capacity gain is more prominent in the region of $0.2 \leq p,q \leq 1$ portrayed by the dark red color area. The maximum gain observed is $G = 1$, which is attained for the values of $(p,q)$ satisfying $pq = 0.5$ for $p,q \leq 1$. We are also interested in the region where we can observe the violation of the bottleneck capacity as well as the associated magnitude. In Fig.~\ref{fig:capacity-bit-phase-flip-pq}(b), we observe that the violation of bottleneck capacity occurs in the region of $p + q \geq 1$ for $p,q \leq 1$. The maximum violation observed is $V = 1$, which is attained for the values of $(p,q) = (0.5,1)$ and $(p,q) = (1,0.5)$. This can be verified directly because the bottleneck capacities for $(p,q) = (0.5,1)$ and $(p,q) = (1,0.5)$ are $C_{\text{E,B}} = 1$. By contrast, for these given values $p$ and $q$, we have $C_{\text{E,Q}} = 2$, which means a perfect entanglement-assisted classical communication over quantum trajectory.

\subsubsection{Quantum Entanglement-Breaking Channels}
\label{Quantum Entanglement-Breaking Channels}

Any quantum channel $\mathcal{N}(\cdot)$ is said to be entanglement breaking if it transforms a maximally-entangled quantum state $\rho_{AB}$, for example $\ket{\Phi^+}_{AB} \bra{\Phi^+}_{AB}$, into a mixture of two separable quantum states $\rho_{A}$ and $\rho_{B}$. Formally, a quantum entanglement-breaking channel is defined as follows~\cite{wilde2013quantum}:
\begin{equation}
(\mathcal{N}_A \otimes I_B)(\rho_{AB}) = \sum_i p_i \left( \rho_{A,i} \otimes \rho_{B,i} \right).
\end{equation} 
An example of quantum entanglement-breaking channels that can be written as a linear combination of Pauli matrices is given by
\begin{equation}
\mathcal{N}(\rho) = \frac{1}{2} \left( X \rho X + Y \rho Y \right).
\label{eq:breaking}
\end{equation}

Now, let us consider two partially entanglement-breaking channels, where the first quantum channel is formulated as
\begin{equation}
\mathcal{D}(\rho) = (1-p) X\rho X + p Y\rho Y,
\label{eq:kraus-breaking-1}
\end{equation}
where the Kraus operators are given by $D_1 = \sqrt{1-p}X$ and $D_2 = \sqrt{p}Y$, where $p = 1/2$ gives us the quantum entanglement-breaking channel of~\eqref{eq:breaking}. Similarly, the second quantum channel is formulated as
\begin{equation}
\mathcal{E}(\rho) = (1-q) X\rho X + q Y\rho Y,
\label{eq:kraus-breaking-2}
\end{equation}
where the Kraus operators are given by $E_1 = \sqrt{1-q}X$ and $E_2 = \sqrt{q}Y$. Based on the Kraus operators of the partially entanglement-breaking channels of~\eqref{eq:kraus-breaking-1} and~\eqref{eq:kraus-breaking-2} and based on the formulation of Kraus operators for classical trajectory of~\eqref{eq:kraus-serial}, we have the following corollary.
\begin{corollary}
\label{col_3}
The entanglement-assisted classical communication capacity $C_{\text{E,C}}$ of two partially entanglement-breaking channels over classical trajectory is given by
\begin{align}
C_{\text{E,C}} &= 2 + \left( 1 - p - q + 2pq \right) \log_2 \left( 1 - p - q + 2pq \right) \nonumber \\
&+ \left( p + q - 2pq \right) \log_2 \left( p + q - 2pq \right).
\label{eq:capacity-serial-breaking}
\end{align}
\begin{IEEEproof}
By substituting $p_1 = 1 - p$, $q_1 = 1 - q$, $p_2 = p$, $q_2 = q$, and the rest of the coefficients equal to $0$ into~\eqref{eq:capacity-serial-general} of Proposition~\ref{prop_1}, we obtain the result in~\eqref{eq:capacity-serial-breaking}. 
\end{IEEEproof}
\end{corollary}
Similarly, based on the formulation of Kraus operators for quantum trajectory of~\eqref{eq:kraus-time-superposition}, we also have the following corollary.
\begin{corollary}
\label{col_4}
The entanglement-assisted classical communication capacity $C_{\text{E,Q}}$ of two partially entanglement-breaking channels over quantum trajectory is given by
\begin{equation}
C_{\text{E,Q}} = 2,
\label{eq:capacity-switch-breaking}
\end{equation}
for every value of $p \in [0,1]$ and $q \in [0,1]$.
\begin{IEEEproof}
By substituting $p_1 = 1 - p$, $q_1 = 1 - q$, $p_2 = p$, $q_2 = q$ and the rest of the coefficients equal to $0$ into~\eqref{eq:capacity-switch-general} of Proposition~\ref{prop_2}, we obtain the result in~\eqref{eq:capacity-switch-breaking}.
\end{IEEEproof}
\end{corollary}

\begin{figure}[t]
\center
\includegraphics[width=\linewidth]{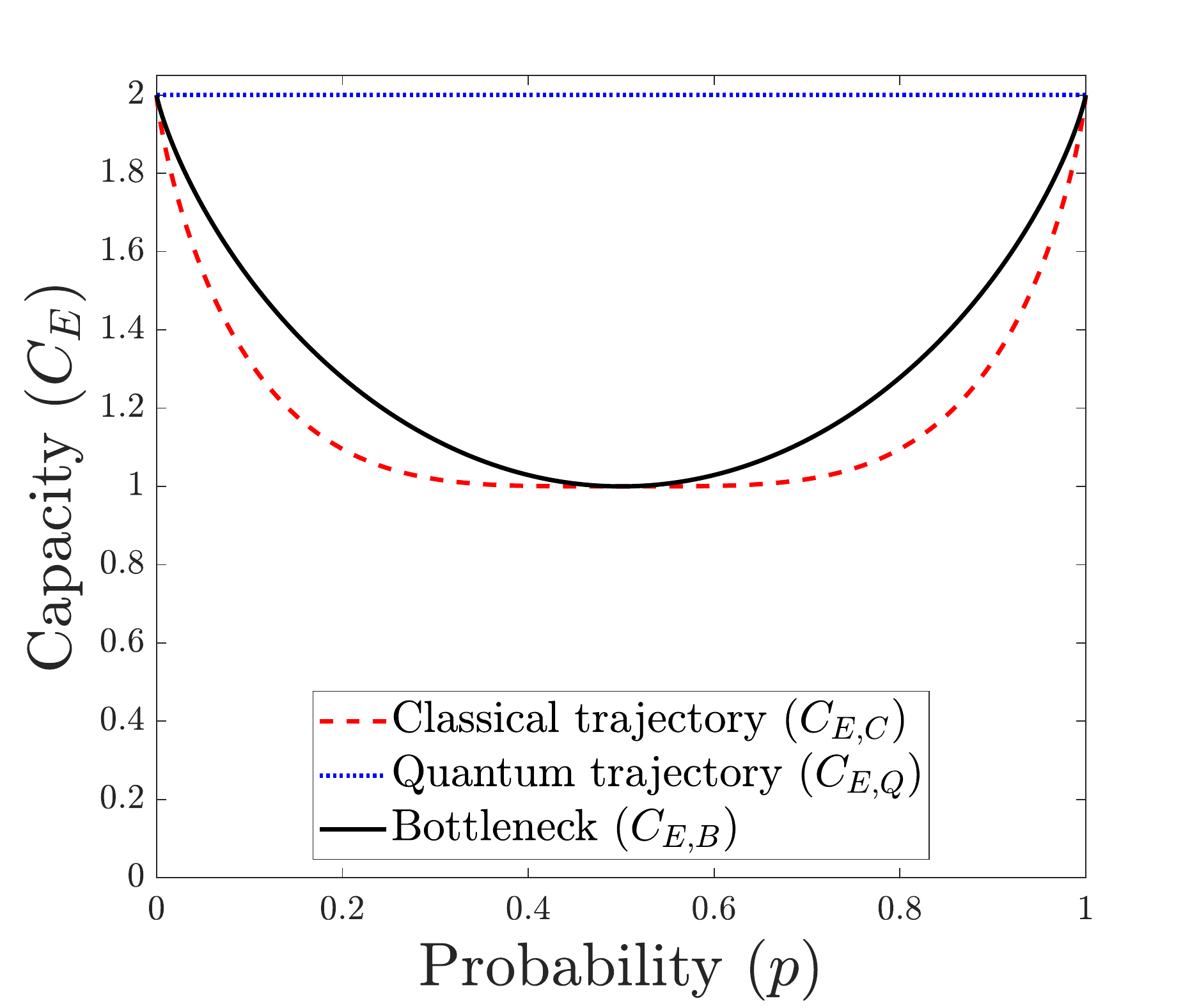}
\caption{The entanglement-assisted classical capacity for two identical partially entanglement-breaking channels. The plots are based on~\eqref{eq:capacity-serial-breaking} and~\eqref{eq:capacity-switch-breaking}.}
\label{fig:capacity-breaking}
\end{figure}

To provide a clearer picture, we portray the entanglement-assisted capacity of~\eqref{eq:capacity-serial-breaking} and~\eqref{eq:capacity-switch-breaking} in a scenario where we have $p = q$ in Fig.~\ref{fig:capacity-breaking}. Once again, we also include the bottleneck capacity of the two partially entanglement-breaking channels. Here, we observe an interesting phenomenon where we can always achieve a perfect entanglement-assisted classical communication over quantum trajectory, which demonstrates a full violation of bottleneck capacity for every value of $p \in (0,1)$. Compared to the result presented in Subsection~\ref{Bit-Flip and Phase-Flip Channels}, the Kraus operators of the partially entanglement-breaking channels contain two non-commuting Pauli matrices given by $XY = -YX$. Therefore, the two different causal orders of two non-commuting Pauli matrices can always be detected by the control qubit $\omega$, since it transforms the initialized $\omega = \ket{+} \bra{+}$ into $\omega = \ket{-} \bra{-}$. By contrast, the remaining combinations of the Pauli matrices given by $XX = YY = I$, leave the control qubit $\omega$ unchanged.

\begin{figure*}[t]
    \centering
    \begin{subfigure}{0.49\linewidth}
        \includegraphics[width=\linewidth]{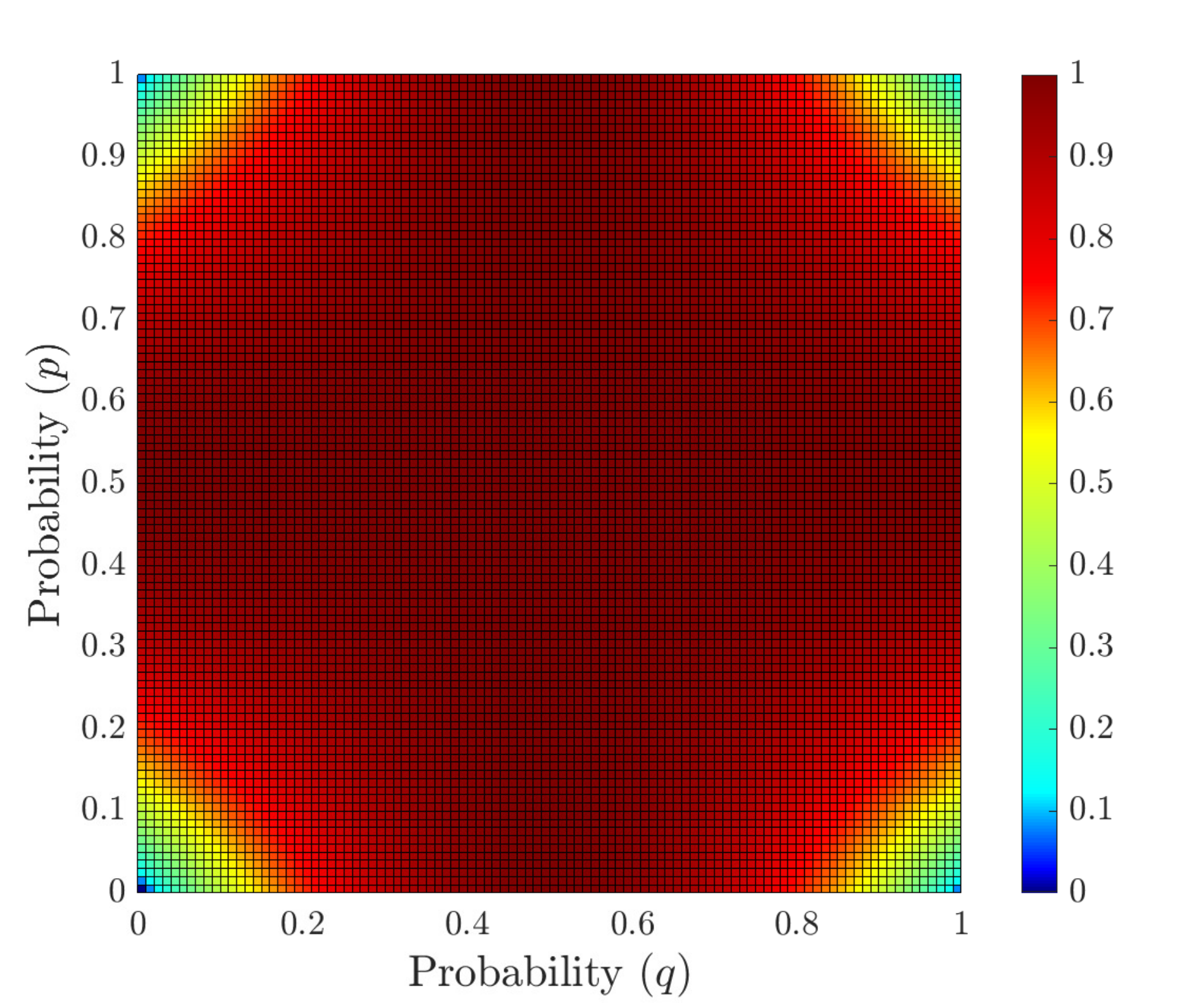}
		\caption{Capacity gain ($G$)}
		\label{fig:capacity-switch-breaking-gain}
    \end{subfigure}
    \hfill
    \begin{subfigure}{0.49\linewidth}
        \includegraphics[width=\linewidth]{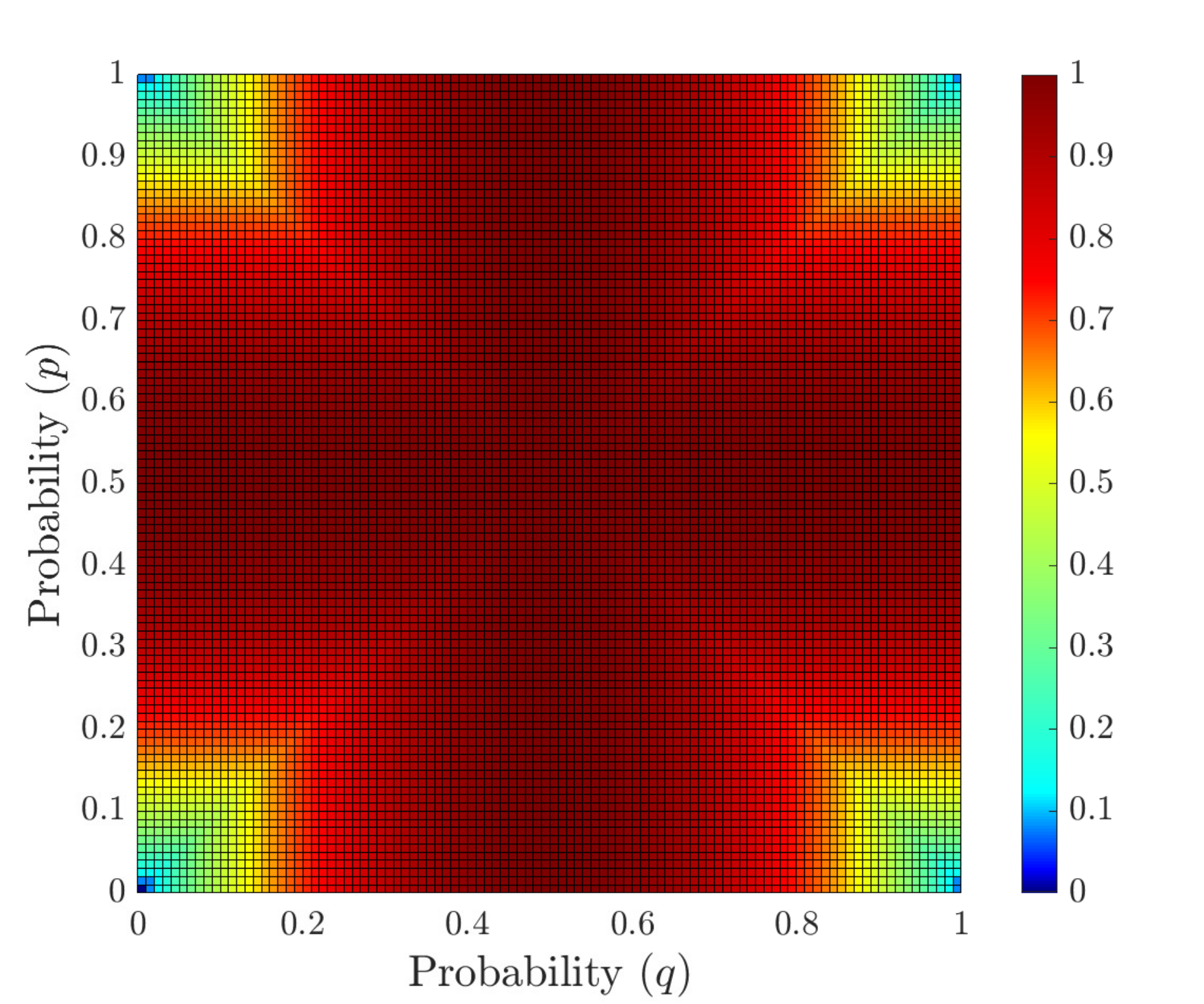}
		\caption{Bottleneck violation ($V$)}
		\label{fig:capacity-switch-breaking-violation}
    \end{subfigure}
\caption{(a) The entanglement-assisted classical capacity gain of partially entanglement-breaking channels over quantum trajectory against classical trajectory. (b) The violation of bottleneck capacity due to the indefinite causal order of quantum channels.}
\label{fig:capacity-breaking-pq}
\hrulefill
\end{figure*}

We portray the capacity gain for two partially entanglement-breaking channels over quantum trajectory in Fig.~\ref{fig:capacity-breaking-pq}(a), while the bottleneck violation in Fig.~\ref{fig:capacity-breaking-pq}(b). Since we always attain a perfect entanglement-assisted classical communication over quantum trajectory for two partially entanglement-breaking channels, the magnitude of capacity gain, which may be observed from the dominantly dark red color area portrayed in Fig.~\ref{fig:capacity-breaking-pq}(a), is significantly higher than that of bit-flip and phase-flip quantum channels. Consequently, the benefit from capacity gain from Fig.~\ref{fig:capacity-breaking-pq}(a) is directly translated to the capacity violation displayed by the color map in Fig.~\ref{fig:capacity-breaking-pq}(b). In this case, we have shown that the quantum trajectory of two entanglement-breaking channels is capable of achieving the full violation of bottleneck capacity for every value $p \in (0,1)$ and $q \in (0,1)$, as described in~\eqref{eq:capacity-switch-breaking} of Corollary~\ref{col_4}.

\begin{remark}
The result in Corollary~\ref{col_4} implies that the indefinite causal order of the quantum channels allows us to always enable a perfect entanglement-assisted classical communication over two partially entanglement-breaking channels.
\end{remark}

\subsubsection{Quantum Depolarizing Channels}
\label{Quantum Depolarizing Channels}

Let us now consider two quantum depolarizing channels. For the first quantum channel, we have the following description:
\begin{equation}
\mathcal{D}(\rho) = (1 - p)\rho + \frac{p}{3} \left( {X}\rho{X} + {Y}\rho{Y} + {Z}\rho{Z} \right),
\label{eq:kraus-depolarizing-1}
\end{equation}
where the Kraus operators are given by $D_1 = \sqrt{1-p}I$, $D_2 = \sqrt{\frac{p}{3}}X$, $D_3 = \sqrt{\frac{p}{3}}Y$, and $D_4 = \sqrt{\frac{p}{3}}Z$. Similarly, for the second quantum channel, we have
\begin{equation}
\mathcal{E}(\rho) = (1 - q)\rho + \frac{q}{3} \left( {X}\rho{X} + {Y}\rho{Y} + {Z}\rho{Z} \right),
\label{eq:kraus-depolarizing-2}
\end{equation}
where the Kraus operators are given by $E_1 = \sqrt{1-q}I$, $E_2 = \sqrt{\frac{q}{3}}X$, $E_3 = \sqrt{\frac{q}{3}}Y$, and $E_4 = \sqrt{\frac{q}{3}}Z$. Based on the Kraus operators of the quantum depolarizing channels of~\eqref{eq:kraus-depolarizing-1} and~\eqref{eq:kraus-depolarizing-2} and based on the formulation of Kraus operators for classical trajectory of~\eqref{eq:kraus-serial}, we arrive at the following corollary.
\begin{corollary}
\label{col_5}
The entanglement-assisted classical communication capacity $C_{\text{E,C}}$ of two quantum depolarizing channels over classical trajectory is given by
\begin{align}
C_{\text{E,C}} &= 2 + \left(1-p-q+\dfrac{4pq}{3}\right) \log_2 \left(1-p-q+\dfrac{4pq}{3}\right) \nonumber \\
&+ \left(p+q-\dfrac{4pq}{3}\right) \log_2 \left( \dfrac{3p+3q-4pq}{9}\right).
\label{eq:capacity-serial-depolarizing}
\end{align}
\begin{IEEEproof}
By substituting $p_0 = q_0 = 1 - p$, $p_1 = p_2 =p_3 = q_1 = q_2 = q_3 = p/3$, into~\eqref{eq:capacity-serial-general} of Proposition~\ref{prop_1}, we obtain the result in~\eqref{eq:capacity-serial-depolarizing}.
\end{IEEEproof}
\end{corollary}
Similarly, based on the the formulation of Kraus operators for quantum trajectory of~\eqref{eq:kraus-time-superposition}, we also arrive at the following corollary.
\begin{corollary}
\label{col_6}
The entanglement-assisted classical communication capacity $C_{\text{E,Q}}$ of two quantum depolarizing channels over quantum trajectory is given by
\begin{align}
C_{\text{E,Q}} &= 2 + H(\alpha) \nonumber \\
&+ \left( 1 - p - q + \frac{4pq}{3} \right) \log_2 \left( 1 - p - q + \frac{4pq}{3} \right) \nonumber \\
&+ (p + q - 2pq) \log_2 \left(\dfrac{p + q - 2pq}{3} \right) + \frac{2pq}{3} \log_2 \frac{2pq}{9},
\label{eq:capacity-switch-depolarizing}
\end{align}
where $\alpha = p_{\ket{+}} = 1 - {2pq}/{3}$.
\begin{IEEEproof}
By substituting $p_0 = 1 - p$, $q_0 = 1 - q$, $p_1 = p_2 = p_3 = p/3$, $q_1 = q_2 = q_3 = q/3$, into~\eqref{eq:capacity-switch-general} of Proposition~\ref{prop_2}, we obtain the result in~\eqref{eq:capacity-switch-depolarizing}.
\end{IEEEproof}
\end{corollary}

\begin{figure}[t]
\center
\includegraphics[width=\linewidth]{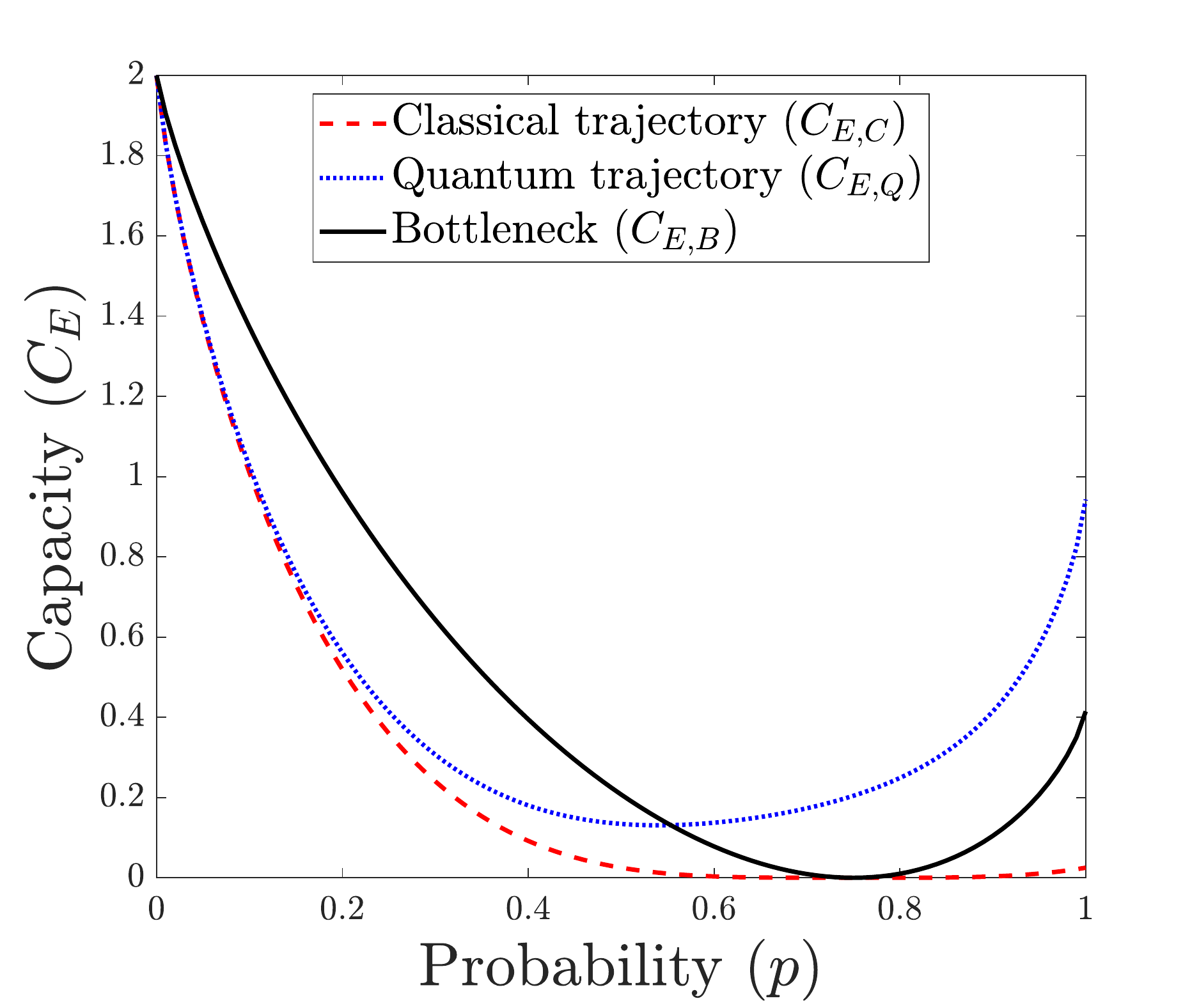}
\caption{The entanglement-assisted classical communication capacity for two identical quantum depolarizing channels. The plots are based on~\eqref{eq:capacity-serial-depolarizing} and~\eqref{eq:capacity-switch-depolarizing}.}
\label{fig:capacity-depolarizing}
\end{figure}

As a special case, we portray the entanglement-assisted capacity of~\eqref{eq:capacity-serial-depolarizing} and~\eqref{eq:capacity-switch-depolarizing} in a scenario where we have $p = q$ in Fig.~\ref{fig:capacity-depolarizing}. In Subsection~\ref{Bit-Flip and Phase-Flip Channels} and~\ref{Quantum Entanglement-Breaking Channels}, the bottleneck capacities are always strictly positive. Instead, here we have a condition where we have a zero-capacity at $p = 0.75$. At this point, we have the so-called fully-depolarizing quantum channel, which we have mentioned briefly in Section~\ref{The Capacity of Entanglement-Assisted Classical Communication}. It means that no classical information can be sent by the means of entanglement-assisted classical communication through the individual quantum channel. Clearly, this is also the case for the classical trajectory since it cannot violate the bottleneck capacity. Interestingly, observe in Fig.~\ref{fig:capacity-depolarizing} that we have a non-zero capacity for entanglement-assisted communication utilizing two fully-depolarizing quantum channels over quantum trajectory. More specifically, we have $C_{\text{E,Q}} = 0.204$ for $p = 0.75$. The ability to enable a non-zero capacity communication using two zero-capacity quantum channels over quantum trajectory is referred to as the \textit{causal activation}~\cite{ebler2018enhanced, salek2018quantum, chiribella2021indefinite, kristjansson2020resource}.

\begin{figure*}[t]
    \centering
    \begin{subfigure}{0.49\linewidth}
        \includegraphics[width=\linewidth]{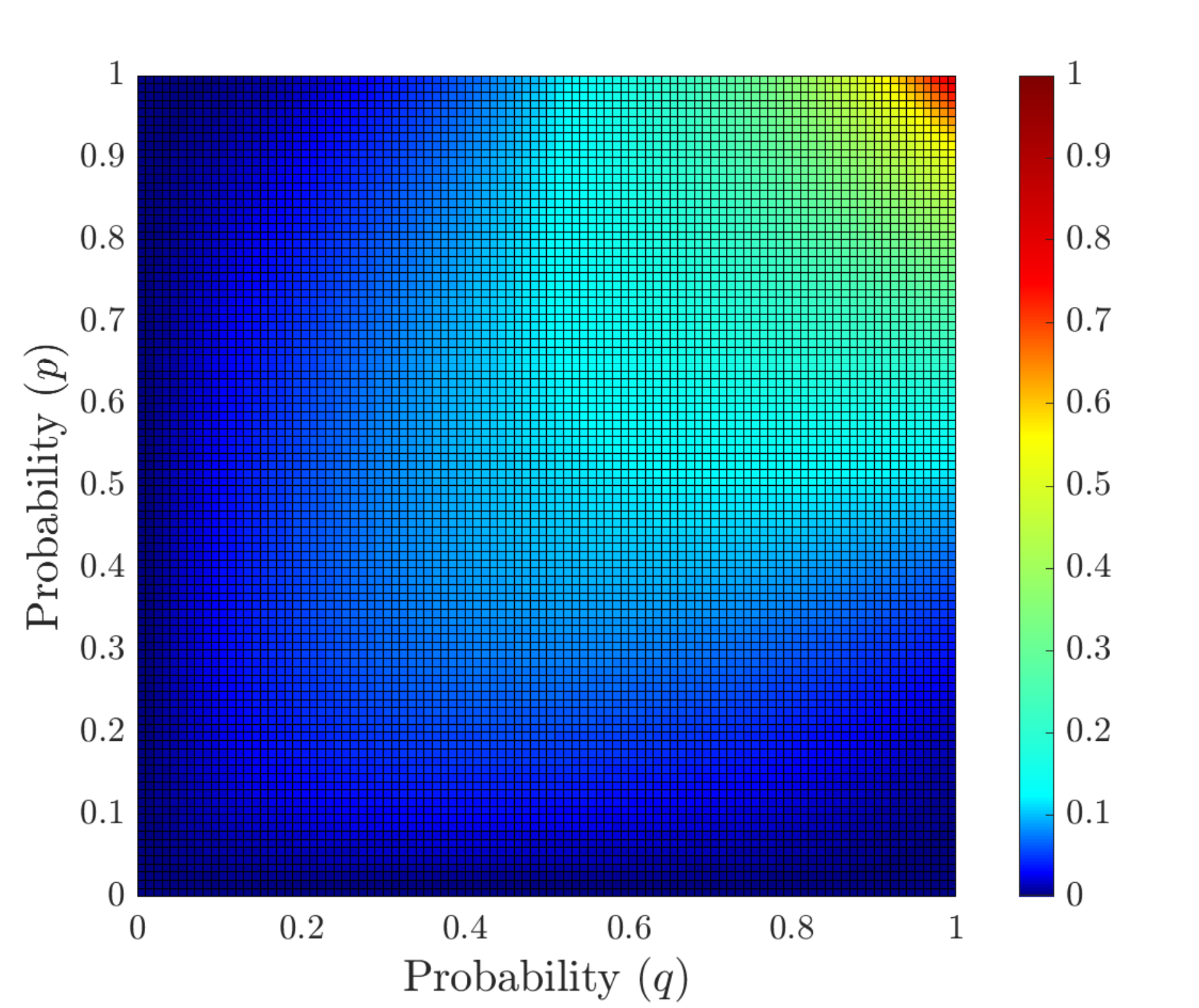}
		\caption{Capacity gain ($G$)}
		\label{fig:capacity-switch-depolarizing-gain}
    \end{subfigure}
    \hfill
    \begin{subfigure}{0.49\linewidth}
        \includegraphics[width=\linewidth]{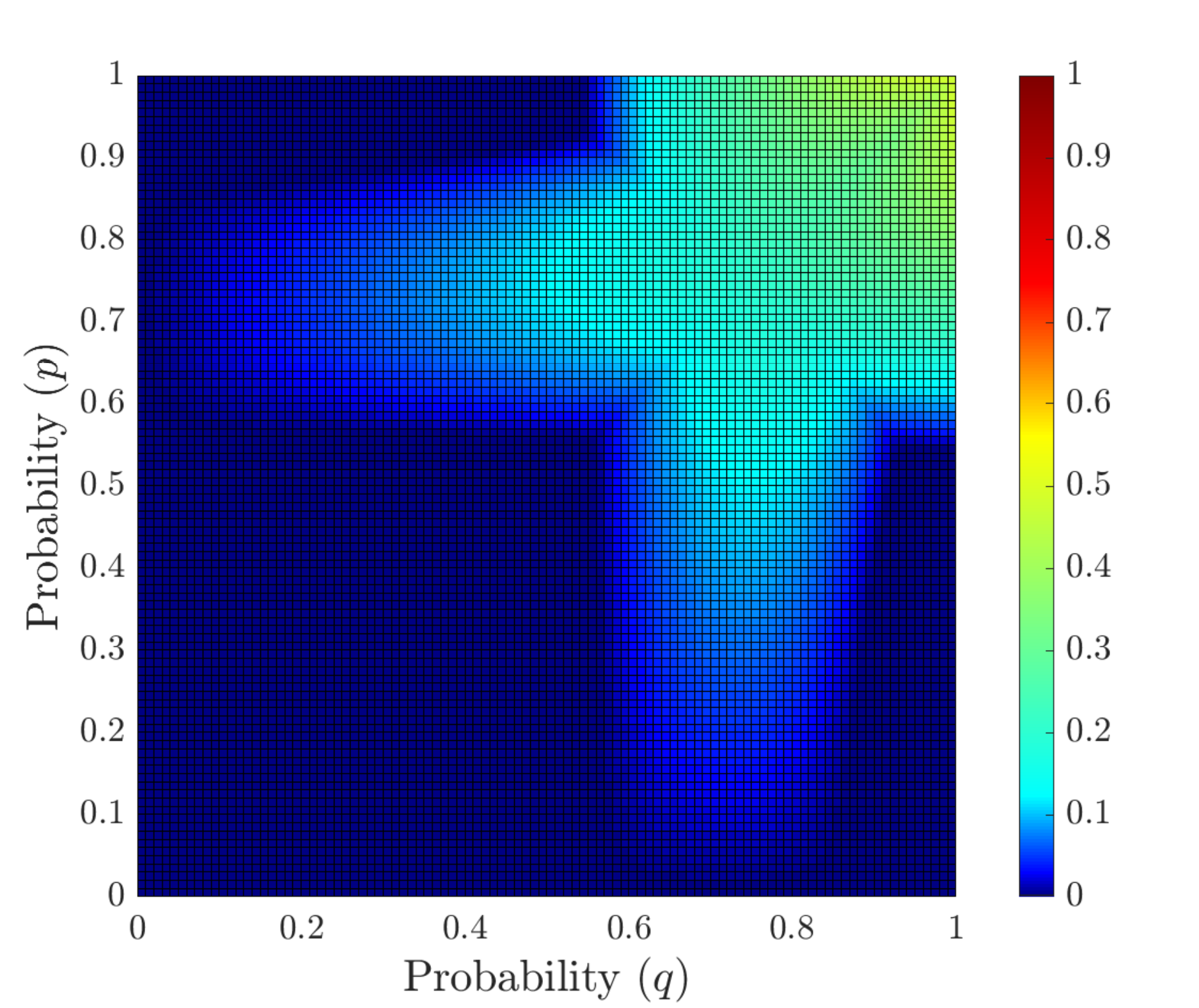}
		\caption{Bottleneck violation ($V$)}
		\label{fig:capacity-switch-depolarizing-violation}
    \end{subfigure}
\caption{(a) The entanglement-assisted classical capacity gain of two quantum depolarizing channels over quantum trajectory against classical trajectory. (b) The violation of bottleneck capacity due to the indefinite causal order of quantum channels.}
\label{fig:capacity-depolarizing-pq}
\hrulefill
\end{figure*}

\begin{remark}
The entanglement-assisted classical communication capacity over quantum trajectory is always strictly positive, which means that we can always send classical information through quantum channels even when we cannot send any classical information through the individual quantum channel. 
\end{remark}

We portray the capacity gain for two quantum depolarizing channels over quantum trajectory in Fig.~\ref{fig:capacity-depolarizing-pq}(a), while the bottleneck violation in Fig.~\ref{fig:capacity-depolarizing-pq}(b). Observe that based on the color map in Fig.~\ref{fig:capacity-depolarizing-pq}(a), the capacity gain gleaned at range values of $0 < p,q < 1$ are relatively modest compared to that of bit-flip and phase-flip quantum channels. The capacity gain are more prominent in the region of $0.5 \leq p,q \leq 1$ denoted by the light blue to dark red color area. The maximum gain observed is $G = 0.918$, which is attained for the value of $(p,q) = (1,1)$. In Fig.~\ref{fig:capacity-depolarizing-pq}(b), we also observe modest violation of bottleneck capacity. Nonetheless, the bottleneck violation can be observed for a wide range of $p$ and $q$ values. The maximum violation observed is $V = 0.528$, which is also attained for the value of $(p,q) = (1,1)$.

\begin{remark}
The quantum trajectory, which leads to indefinite causal order of quantum channels, enables the violation of bottleneck capacity, implying that the phenomenon inducing this violation cannot be obtained by any process exhibiting a well-defined causal order.
\end{remark}

\subsection{Quantum Communication Capacity}
\label{Quantum Communication Capacity}

In the previous subsections, we have provided a thorough analysis on the entanglement-assisted classical communication capacity via classical and quantum trajectories. In this section, we extend the analysis to the entanglement-assisted quantum communication. In~\cite{bennett1999entanglement} and~\cite{bowen2002entanglement}, the relationship between the entanglement-assisted capacity of classical communication $(C_E)$ and that of quantum capacity $(Q_E)$ is readily given by
\begin{equation}
    Q_E = \frac{C_E}{2},
	\label{eq:classical-to-quantum}
\end{equation}
based on quantum superdense coding versus quantum teleportation trade-off, which dictates that a pair of pre-shared maximally-entangled quantum state and a single use of quantum channel $\mathcal{N}(\cdot)$ can be exchanged for a single qubit or two classical bits. Consequently, by exploiting~\eqref{eq:classical-to-quantum}, the results in Proposition~\ref{prop_1} and~\ref{prop_2} for entanglement-assisted classical communication over classical and quantum trajectories can be extended directly to the entanglement-assisted quantum communications as shown in the following corollaries.

\begin{corollary}
\label{col_7}
The entanglement-assisted quantum communication capacity $Q_{\text{E,Q}}$ of the two arbitrary quantum Pauli channels of~\eqref{eq:channeld} and~\eqref{eq:channele} over classical trajectory is given by
\begin{align}
Q_{\text{E,C}} &= \dfrac{C_{\text{E,C}}}{2} = 1 + \frac{1}{2} \left[ {A}_0 \log_2 {A}_0 + {A}_1 \log_2 {A}_1 \right. \nonumber \\
&+ \left. {A}_2 \log_2 {A}_2 + {A}_3 \log_2 {A}_3 \right].
\label{eq:capacity-switch-bit-phase-classical}
\end{align}
\begin{IEEEproof}
The proof follows directly by accounting for Proposition~\ref{prop_1} and \eqref{eq:classical-to-quantum}.
\end{IEEEproof}
\end{corollary}

\begin{corollary}
\label{col_8}
The entanglement-assisted quantum communication capacity $Q_{\text{E,Q}}$ of the two arbitrary quantum Pauli channels of~\eqref{eq:channeld} and~\eqref{eq:channele} over quantum trajectory is given by
\begin{align}
Q_{\text{E,Q}} = \dfrac{C_{\text{E,Q}}}{2} &= 1 + \frac{1}{2} \left[ H(\alpha) + {A}_0 \log_2 {A}_0 + {A}_1^+ \log_2 {A}_1^+ \right. \nonumber \\
&+ \left. {A}_2^+ \log_2 {A}_2^+ + {A}_3^+ \log_2 {A}_3^+ + {A}_1^- \log_2 {A}_1^- \right. \nonumber \\
&+ \left. {A}_2^- \log_2 {A}_2^- + {A}_3^- \log_2 {A}_3^- \right].
\label{eq:capacity-switch-bit-phase-quantum}
\end{align}
\begin{IEEEproof}
The proof follows directly by accounting for Proposition~\ref{prop_2} and \eqref{eq:classical-to-quantum}.
\end{IEEEproof}
\end{corollary}

It may be interesting to see the relationship between the capacity bounds of quantum communication via quantum trajectory presented in~\cite{cacciapuoti2019capacity} to the results presented here. Let us take the example of bit-flip and phase flip quantum channels of~\eqref{eq:kraus-bit-flip} and~\eqref{eq:kraus-phase-flip}. Based on~\eqref{eq:capacity-switch-bit-phase-quantum} in Corollary~\ref{col_8}, we obtain the entanglement-assisted quantum communication capacity $Q_{\text{E,Q}}$ of the bit-flip and phase-flip quantum channels over quantum trajectory as follows:
\begin{align}
Q_{\text{E,Q}} &= 1 + \dfrac{1}{2} \left[ H(\alpha) + (1-p)^2 \log_2 (1-p)^2 \right. \nonumber \\
&+ \left. (2p-2p^2) \log_2 (p-p^2) + p^2 \log_2 p^2 \right],
\label{eq:quantum-capacity-switch-bit-phase}
\end{align}
for $p = q$, where $\alpha = 1 - p^2$. Meanwhile, the lower bound of the quantum communication capacity over quantum trajectory is given by~\cite{cacciapuoti2019capacity}
\begin{equation}
Q_{LB} = p^2 + \max \lbrace 0, 1 - p^2 - 2H(p) + H(p^2) \rbrace.
\label{eq:quantum-lower}
\end{equation}
This lower bound is obtained from the entropy measure of the quantum Pauli channels, which represents the capacity of unassisted quantum communication~\cite{lloyd1997capacity, devetak2005private}. By contrast, quantum communication capacity over quantum trajectory is upper-bounded by the capacity of two-way entanglement-assisted quantum communication~\cite{pirandola2017fundamental}. In case of bit-flip and phase-flip quantum channels, this upper bound is given by~\cite{cacciapuoti2019capacity}
\begin{equation}
Q_{UB} = 1 - \left[ (1-p)H(p) \right],
\label{eq:quantum-upper}
\end{equation}
which is obtained from the relative entropy of entanglement of the Choi matrix~\cite{pirandola2017fundamental}.

\begin{figure}[ht!]
\center
\includegraphics[width=\linewidth]{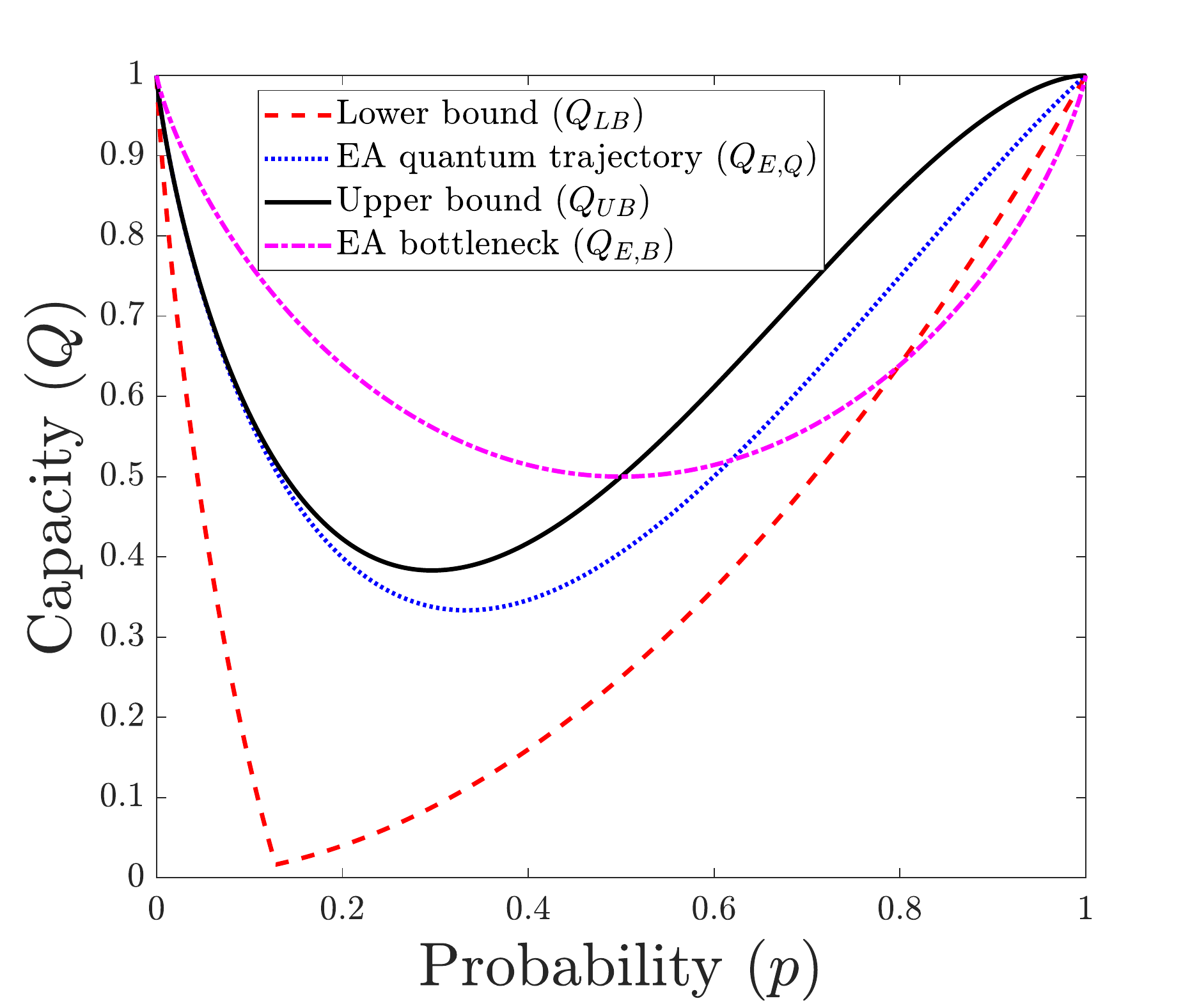}
\caption{The capacity of quantum communication for the combination of bit-flip and phase-flip quantum channels.}
\label{fig:capacity-bit-phase-flip-quantum}
\end{figure}

We plot the entanglement-assisted capacity of~\eqref{eq:quantum-capacity-switch-bit-phase} and the corresponding quantum communication capacity bounds of~\eqref{eq:quantum-lower} and~\eqref{eq:quantum-upper} in Fig.~\ref{fig:capacity-bit-phase-flip-quantum}. Again, we include the bottleneck entanglement-assisted quantum communication capacity of the bit-flip and phase-flip quantum channels, which is given by
\begin{equation}
Q_{\text{E,B}} = 1 - \dfrac{H(p)}{2},
\label{eq:quantum-bottleneck}
\end{equation}
for $p = q$. The result presented in Fig.~\ref{fig:capacity-bit-phase-flip-quantum} suggests the confirmation of the following relationship between various capacities of quantum communication over quantum trajectory: $Q_{LB} \leq Q_{\text{E,Q}} \leq Q_{UB}$. Additionally, the violation of bottleneck capacity for entanglement-assisted quantum communication can be observed within the range of $0.618 < p < 1$.

\begin{figure*}[t]
    \centering
    \begin{subfigure}{0.49\linewidth}
        \includegraphics[width=\linewidth]{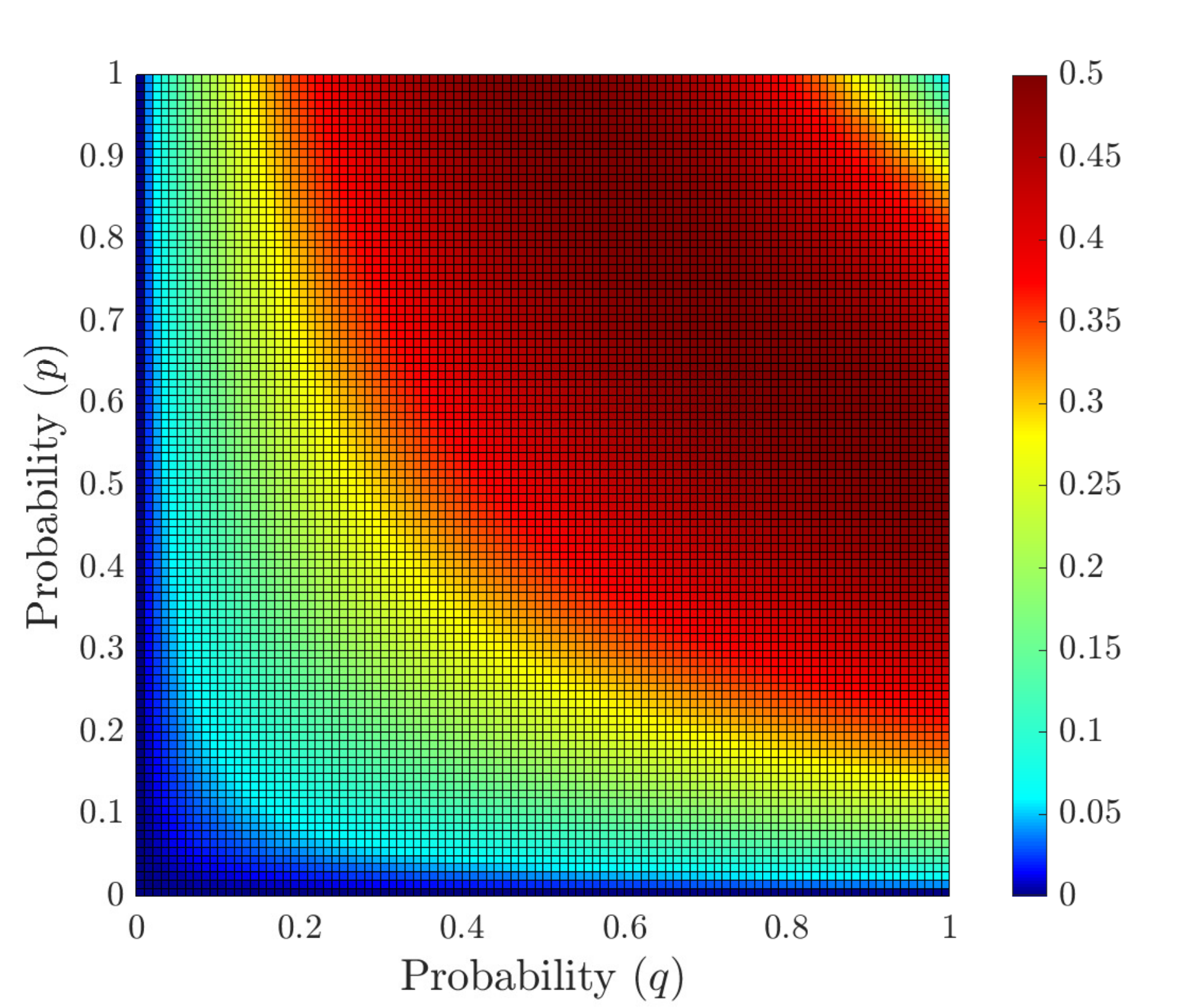}
		\caption{Capacity gain ($G$)}
		\label{fig:capacity-switch-bit-phase-gain-quantum}
    \end{subfigure}
    \hfill
    \begin{subfigure}{0.49\linewidth}
        \includegraphics[width=\linewidth]{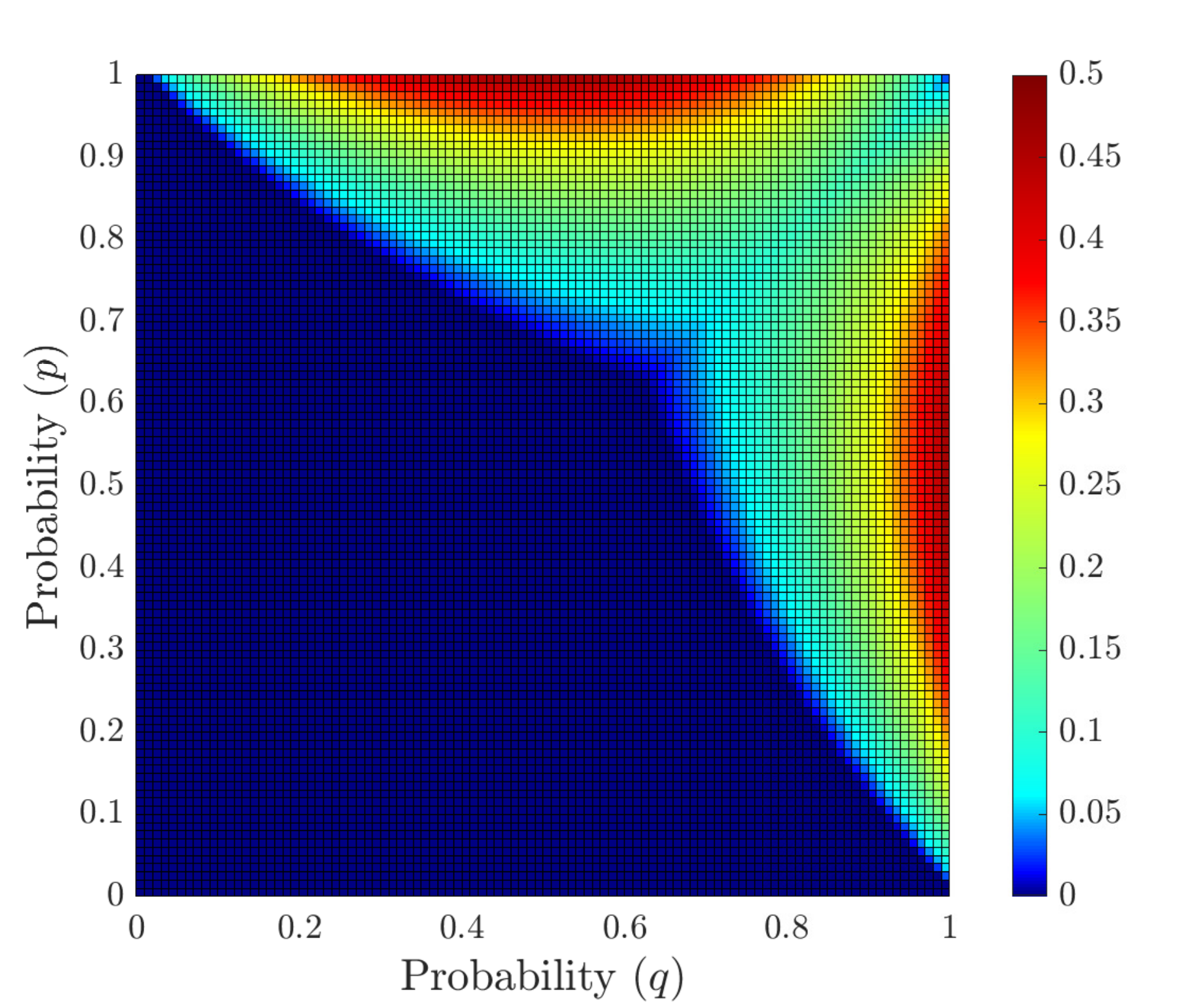}
		\caption{Bottleneck violation ($V$)}
		\label{fig:capacity-switch-bit-phase-violation-quantum}
    \end{subfigure}
\caption{(a) The entanglement-assisted quantum capacity gain of the bit-flip and phase-flip quantum channels over quantum trajectory against classical trajectory. (b) The violation of bottleneck capacity due to the indefinite causal order of quantum channels.}
\label{fig:capacity-bit-phase-pq-quantum}
\hrulefill
\end{figure*}

Following the same line of investigation presented for entanglement-assisted classical communication, here we also provide the results of capacity gain and bottleneck violation of entanglement-assisted quantum communication over quantum trajectory. More specifically, let us observe Fig.~\ref{fig:capacity-bit-phase-pq-quantum}. We plot the capacity gain for the bit-flip and phase-flip quantum channels over quantum trajectory in Fig.~\ref{fig:capacity-bit-phase-pq-quantum}(a) and the bottleneck violation in Fig.~\ref{fig:capacity-bit-phase-pq-quantum}(b). We can observe in Fig.~\ref{fig:capacity-bit-phase-pq-quantum}(a) that we obtain the capacity gain at all range values of $0 < p,q < 1$. Similar to the results for entanglement-assisted classical communication, the capacity gain is more profound in the region of $0.2 \leq p,q \leq 1$ portrayed by the dark red color area. The maximum gain observed is $G = 0.5$, which is attained for the values of $(p,q)$ satisfying $pq = 0.5$ for $p,q \leq 1$. Notice that the maximum gain attained in entanglement-assisted quantum communication is half of its classical counterpart due to the trade-off of~\eqref{eq:classical-to-quantum}. In Fig.~\ref{fig:capacity-bit-phase-pq-quantum}(b), we observe that the violation of bottleneck capacity occurs in the region of $p + q \geq 1$ for $p,q \leq 1$ with the maximum violation observed is $V = 0.5$, which is attained for the values of $(p,q) = (0.5,1)$ and $(p,q) = (1,0.5)$. This can be verified directly since the bottleneck capacities for $(p,q) = (0.5,1)$ and $(p,q) = (1,0.5)$ are given by $Q_{\text{E,B}} = 0.5$. Meanwhile, for the values $(p,q) = (0.5,1)$ and $(p,q) = (1,0.5)$, we have $Q_{\text{E,Q}} = 1$, which means we can achieve a perfect entanglement-assisted quantum communication over quantum trajectory.

To conclude this section, we would like to highlight some potential practical advantages that may be gained based on the resultant $G$ and $V$ gleaned from the indefinite causal order of the quantum channels. As an instance, in the area of quantum error-correction codes, the capacity gain may be utilized for improving the quantum coding rate or the quantum bit error ratio of the designed codes for mitigating the quantum errors imposed by the quantum Pauli channels. Another prospective application may be constituted by quantum-secure direct-communication, where the indefinite causal order of the quantum channels can be utilized for increasing the transmission rate. Finally, the resultant capacity gain may also be used for increasing the fidelity of a remote two-qubit quantum gate in a distributed quantum computing framework. However, it is important to note that the application aspect of the quantum configuration resulting in indefinite causal order within the quantum communication framework is still in its infancy and thus, further investigation is mandatory. It is also crucial to underline that the motivation for studying the advantage of the indefinite causal order is not directly a practical one, since such motivation is related to explore how the theory of quantum communication -- as we know it -- would be affected by the possibility of combining quantum channels in a superposition of causal orders, as clarified in~\cite{kristjansson2020resource}.

\section{Conclusions and Future Works}
\label{Conclusions and Future Works}

In this contribution, we have presented the general formulation of both entanglement-assisted classical and quantum communication capacities over quantum trajectory. Our analysis included several examples from the family of quantum Pauli channels. Furthermore, we have explicitly portrayed the region in which the indefinite causal structure of quantum channels violate the bottleneck capacity constraining the capacity of quantum channels with a definite causal structure. Additionally, for two fully-depolarizing quantum channels, we have witnessed the causal activation of the zero capacity quantum channels induced by quantum trajectory.

For our next works, we may consider extending our analysis for more than two quantum channels~\cite{procopio2019communication, procopio2020sending, sazim2021classical, chiribella2020quantum}. In this treatise, the control qubit $\omega$ exhibits only one degree of freedom and therefore only one superposition of two classical trajectories can be included. When the number of quantum channels increases $(N > 2)$, we can no longer use a qubit as our control qubit $\omega$, but a qudit with $d$ quantum level. The critical point arising from this scenario is the ultimate or asymptotic limit of capacity gain can be attained by exploiting indefinite causal order of infinitely many quantum channels. Finally, we are also interested in investigating the effect of the indefinite causal order of quantum channels in a more realistic scenario of quantum superdense coding, where the quantum channels affect both the forward and backward communication between $A$ and $B$~\cite{laurenza2020dense}. 
\appendices

\section{Proof of Proposition 1}
\label{Proof of Proposition Serial}

Let us consider two quantum Pauli channels $\mathcal{D}(\cdot)$ and $\mathcal{E}(\cdot)$. By substituting the Kraus operators description of quantum Pauli channels, into the Kraus operators formulation for classical trajectory given in~\eqref{eq:kraus-serial}, we obtain the resultant quantum channel $\mathcal{S}(\cdot)$ as follows:
\begin{align}
\mathcal{S}(\mathcal{D},\mathcal{E})(\rho) &= (p_0 q_0 + p_1 q_1 + p_2 q_2 + p_3 q_3) \rho \nonumber \\
&+ (p_0 q_1 + p_1 q_0 + p_2 q_3 + p_3 q_2) X\rho X \nonumber \\
&+ (p_0 q_2 + p_2 q_0 + p_3 q_1 + p_1 q_3) Y\rho Y \nonumber \\
&+ (p_0 q_3 + p_3 q_0 + p_1 q_2 + p_2 q_1) Z\rho Z.
\label{eq:resultant-serial-general}
\end{align}
Based on the resultant quantum channel $\mathcal{S}(\cdot)$ of~\eqref{eq:resultant-serial-general}, we obtain the following transition probability:
\begin{equation}
p(y_i|x) = \left\{
\begin{array}{ll}
p_0 q_0 + p_1 q_1 + p_2 q_2 + p_3 q_3, & \text{for} \ i = 0 \\
p_0 q_1 + p_1 q_0 + p_2 q_3 + p_3 q_2, & \text{for} \ i = 1 \\
p_0 q_2 + p_2 q_0 + p_3 q_1 + p_1 q_3, & \text{for} \ i = 2 \\
p_0 q_3 + p_3 q_0 + p_1 q_2 + p_2 q_1, & \text{for} \ i = 3
\end{array}.
\right.
\label{eq:transition-serial-general}
\end{equation}
Substituting the transition probability of~\eqref{eq:transition-serial-general} into~\eqref{eq:capacity-specific-simple}, we obtain the entanglement-assisted classical capacity over classical trajectory of~\eqref{eq:capacity-serial-general} presented in Proposition~\ref{prop_1}. Hence, we complete our proof.

\section{Proof of Proposition 2}
\label{Proof of Proposition Switch}

By substituting the Kraus operators description of quantum Pauli channels, into the Kraus operators formulation for quantum trajectory given in~\eqref{eq:kraus-time-superposition}, we obtain the resultant channel $\mathcal{S}_{\omega}(\cdot)$ as follows:
\begin{align}
\mathcal{S}(\mathcal{D},\mathcal{E})_{\omega}(\rho) &= \big[ (p_0 q_0 + p_1 q_1 + p_2 q_2 + p_3 q_3) \rho \nonumber \\
&+ (p_0 q_1 + p_1 q_0) X\rho X + (p_0 q_2 + p_2 q_0) Y\rho Y \nonumber \\
&+ (p_0 q_3 + p_3 q_0) Z\rho Z \big] \otimes \ket{+} \bra{+} \nonumber \\
&+ \big[(p_2 q_3 + p_3 q_2) X\rho X + (p_3 q_1 + p_1 q_3) Y\rho Y \nonumber \\ 
&+ (p_1 q_2 + p_2 q_1) Z\rho Z \big] \otimes \ket{-} \bra{-}.
\label{eq:resultant-switch-general}
\end{align}
Observe that we may measure the control qubit $\omega$ in the Hadamard basis $\lbrace \ket{+} \bra{+}, \ket{-} \bra{-} \rbrace$. The measurement of the control qubit $\omega$ collapses the resultant channel $\mathcal{S}_{\omega}(\cdot)$ into either $\mathcal{S}_{\ket{+}}(\cdot)$ or $\mathcal{S}_{\ket{-}}(\cdot)$ depending on the result. Let us denote $p_{\ket{+}}$ as the probability of measuring the control qubit in $\ket{+} \bra{+}$ state and $p_{\ket{-}}$ as the probability of measuring the control qubit in $\ket{-} \bra{-}$ state. Therefore, the entanglement-assisted classical communication capacity $C_{\text{E,Q}}$ over quantum trajectory can be generalized as
\begin{equation}
C_{\text{E,Q}} = p_{\ket{+}}C_E(\mathcal{S}_{\ket{+}}) + p_{\ket{-}}C_E(\mathcal{S}_{\ket{-}}),
\label{eq:capacity-switch}
\end{equation}
where $C_E(\mathcal{S}_{\ket{+}})$ and $C_E(\mathcal{S}_{\ket{-}})$ are the capacity of quantum channel $\mathcal{S}_{\ket{+}}(\cdot)$ and quantum channel $\mathcal{S}_{\ket{-}}(\cdot)$, respectively. Explicitly, based on~\eqref{eq:resultant-switch-general}, we measure $\omega = \ket{+} \bra{+}$ with a probability of $p_{\ket{+}} = p_0 q_0 + p_1 q_1 + p_2 q_2 + p_3 q_3 + p_0 q_1 + p_1 q_0 + p_0 q_2 + p_2 q_0 + p_0 q_3 + p_3 q_0$, which consequently collapses the resultant channel $\mathcal{S}_{\omega}(\cdot)$ to
\begin{align}
\mathcal{S}_{\ket{+}}(\mathcal{D},\mathcal{E})(\rho) &= \dfrac{1}{p_{\ket{+}}} \big[ (p_0 q_0 + p_1 q_1 + p_2 q_2 + p_3 q_3) \rho \nonumber \\ 
&+ (p_0 q_1 + p_1 q_0) X\rho X + (p_0 q_2 + p_2 q_0) Y\rho Y \nonumber \\
&+ (p_0 q_3 + p_3 q_0) Z\rho Z \big].
\label{eq:resultant-switch-general-plus}
\end{align}
By contrast, we measure $\omega = \ket{-} \bra{-}$ with a probability of $p_{\ket{-}} = p_2 q_3 + p_3 q_2 + p_3 q_1 + p_1 q_3 + p_1 q_2 + p_2 q_1$, which consequently collapses the resultant channel $\mathcal{S}_{\omega}(\cdot)$ to
\begin{align}
\mathcal{S}_{\ket{-}}(\mathcal{D},\mathcal{E})(\rho) &= \dfrac{1}{p_{\ket{-}}} \big[ (p_2 q_3 + p_3 q_2) X\rho X 
\nonumber \\
&+ (p_3 q_1 + p_1 q_3) Y\rho Y + (p_1 q_2 + p_2 q_1) Z\rho Z \big].
\label{eq:resultant-switch-general-minus}
\end{align}
Based on~\eqref{eq:resultant-switch-general-plus} and~\eqref{eq:resultant-switch-general-minus}, we can devise the transition probability $p(y_i|x)$ and determine the capacity of $C_E(\mathcal{S}_{\ket{+}})$ and $C_E(\mathcal{S}_{\ket{-}})$ accordingly. Substituting them into~\eqref{eq:capacity-switch}, we obtain the entanglement-assisted classical communication capacity of quantum two quantum Pauli channels over quantum trajectory of~\eqref{eq:capacity-switch-general} presented in Proposition~\ref{prop_2}. Hence, we complete our proof.

\bibliographystyle{ieeetr}

\begin{IEEEbiography}[{\includegraphics[width=1in,height=1.25in,clip,keepaspectratio]{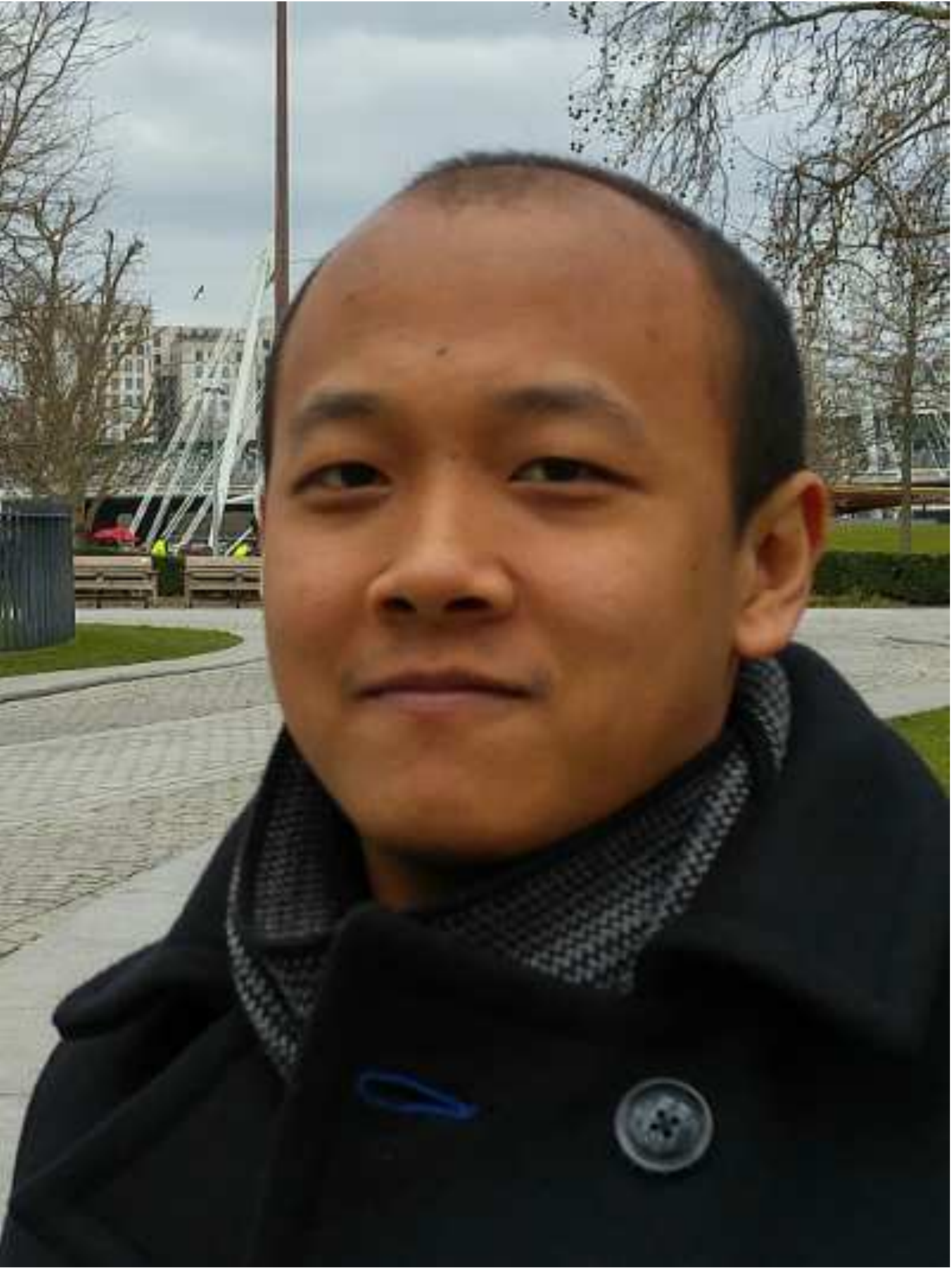}}]{Daryus Chandra} (S'15, M'20) received the M.Eng. degree in electrical engineering from Universitas Gadjah Mada, Indonesia, in 2014 and the Ph.D. degree in electronics and electrical engineering from University of Southampton, UK, in 2020. He was a research fellow with the Future Communications Laboratory, University of Naples Federico II, Italy. Currently, he is a research fellow with the Next-Generation Wireless Research Group, University of Southampton, UK. His research interests include classical and quantum error correction codes, quantum information, and quantum communications.
\end{IEEEbiography}

\begin{IEEEbiography}
[{\includegraphics[width=1in,height=1.25in,clip,keepaspectratio]{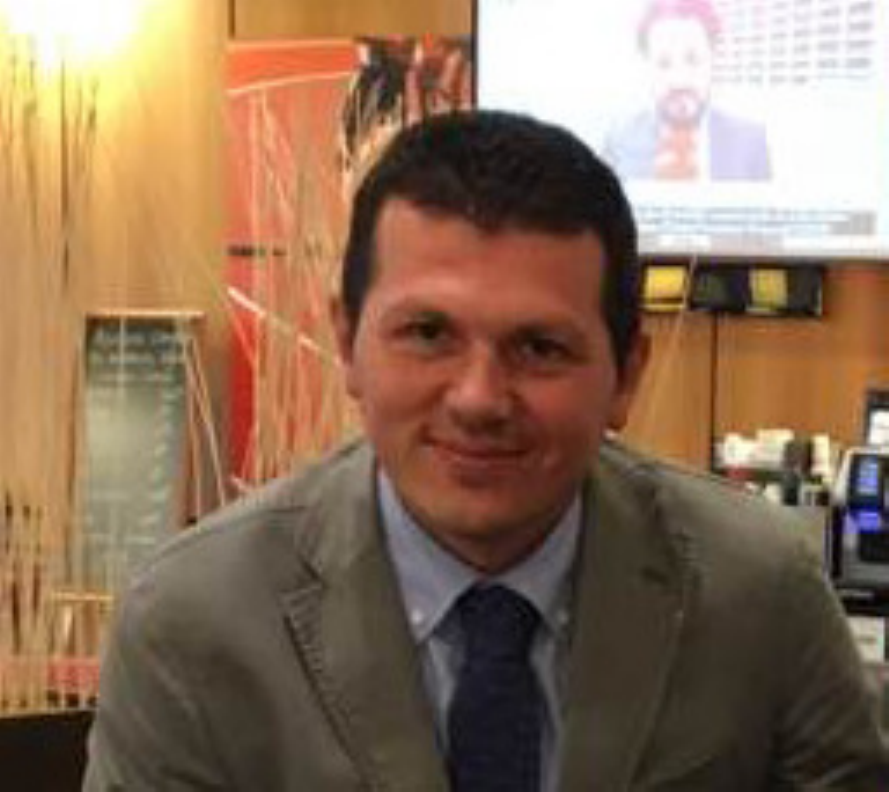}}]{Marcello Caleffi} (M'12, SM'16) received the M.S. degree with the highest score (summa cum laude) in computer science engineering from the University of Lecce, Lecce, Italy, in 2005, and the Ph.D. degree in electronic and telecommunications engineering from the University of Naples Federico II, Naples, Italy, in 2009.  Currently, he is associate professor with the DIETI Department, University of Naples Federico II. From 2010 to 2011, he was with the Broadband Wireless Networking Laboratory at Georgia Institute of Technology, Atlanta, as visiting researcher. In 2011, he was also with the NaNoNetworking Center in Catalunya (N3Cat) at the Universitat Politecnica de Catalunya (UPC), Barcelona, as visiting researcher. Since July 2018, he held the Italian national habilitation as \textit{Full Professor} in Telecommunications Engineering. His work appeared in several premier IEEE Transactions and Journals, and he received multiple awards, including \textit{best strategy} award, \textit{most downloaded article} awards and \textit{most cited article} awards. Currently, he serves as \textit{associate editor/associate technical editor} for IEEE Communications Magazine, IEEE Trans. on Wireless Communications, IEEE Trans. on Quantum Engineering and IEEE Communications Letters. He served as Chair, TPC Chair, Session Chair, and TPC Member for several premier IEEE conferences. In 2016, he was elevated to IEEE Senior Member and in 2017 he has been appointed as Distinguished Lecturer from the \textit{IEEE Computer Society}. In December 2017, he has been elected Treasurer of the Joint \textit{IEEE VT/ComSoc Chapter Italy Section}. In December 2018, he has been appointed member of the IEEE \textit{New Initiatives Committee}.
\end{IEEEbiography}

\begin{IEEEbiography}
[{\includegraphics[width=1in,height=1.25in,clip,keepaspectratio]{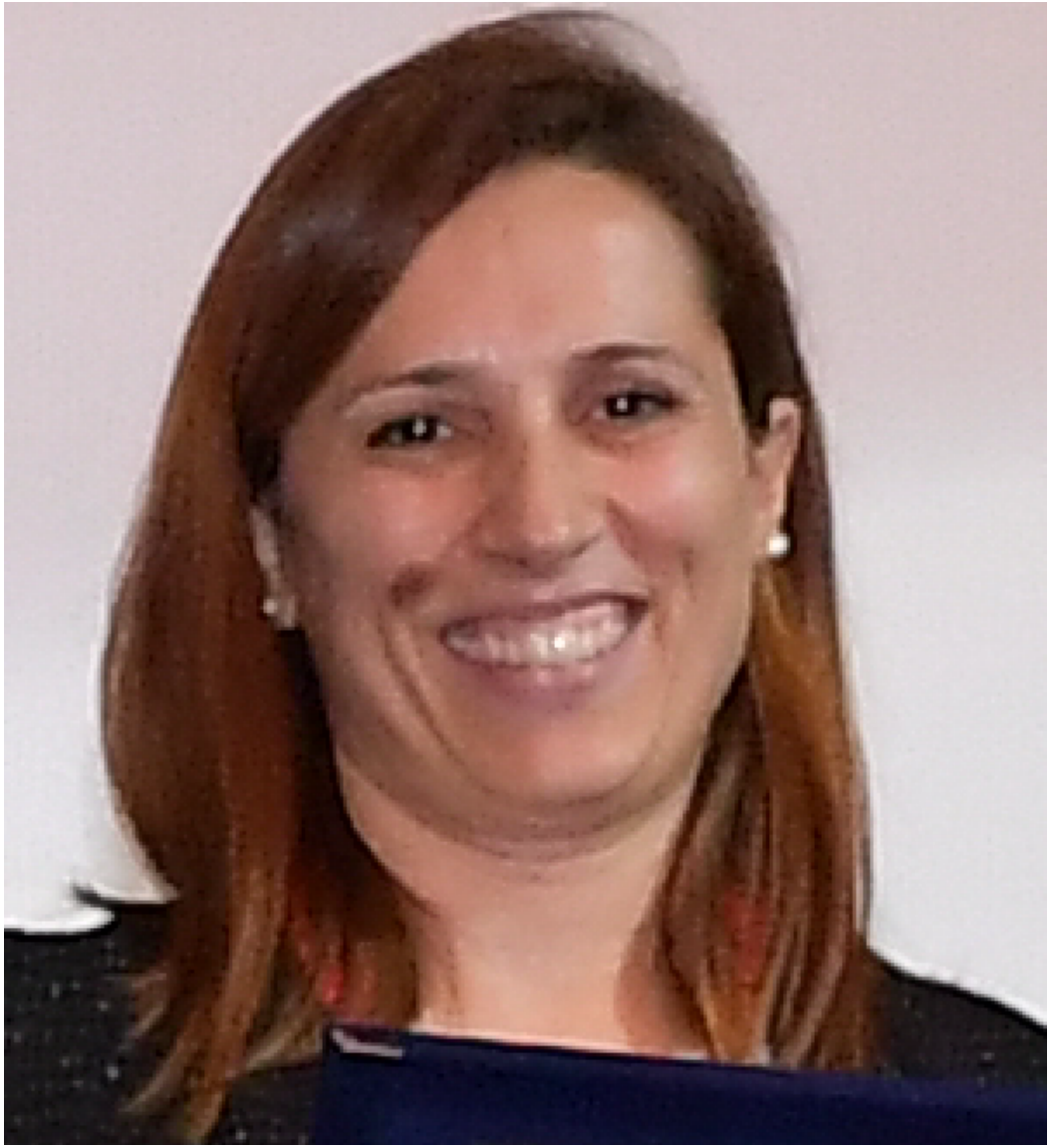}}]{Angela Sara Cacciapuoti} (M'10, SM'16) is an Associate Professor at the University of Naples Federico II, Italy. Since July 2018, she held the national habilitation as ``Full Professor" in Telecommunications Engineering. Her work has appeared in first tier IEEE journals and she has received different awards, including the elevation to the grade of IEEE Senior Member in 2016. Currently, Angela Sara serves as \textit{Area Editor} for IEEE Communications Letters, and as \textit{Editor/Associate Editor} for the journals: IEEE Trans. on Communications, IEEE Trans. on Wireless Communications, IEEE Trans. on Quantum Engineering, IEEE Network and IEEE Open Journal of Communications Society. She was a recipient of the 2017 Exemplary Editor Award of the IEEE Communications Letters. In 2016 she has been an appointed member of the IEEE ComSoc Young Professionals Standing Committee. From 2017 to 2018, she has been the Award Co-Chair of the N2Women Board. From 2017 to 2020, she has been the Treasurer of the IEEE Women in Engineering (WIE) Affinity Group of the IEEE Italy Section. In 2018, she has been appointed as Publicity Chair of the IEEE ComSoc Women in Communications Engineering (WICE) Standing Committee. And since 2020, she is the vice-chair of WICE. Her current research interests are mainly in Quantum Communications, Quantum Networks and Quantum Information Processing.
\end{IEEEbiography}

\end{document}